\documentclass[conference]{IEEEtran}
\IEEEoverridecommandlockouts
\usepackage{cite}
\usepackage{amsmath,amssymb,amsfonts}
\usepackage{algorithm,algorithmicx}
\usepackage{algpseudocode}
\usepackage{graphicx}
\usepackage{textcomp}
\usepackage{xcolor}

\usepackage{subfigure}
\usepackage{array}
\usepackage{pdfpages}
\usepackage{tabularx}
\usepackage{stfloats}
\usepackage{url}
\usepackage{verbatim}
\usepackage{mathrsfs}
\usepackage{cite}
\usepackage{caption}
\usepackage{float} 
\usepackage{subcaption}
\usepackage{grffile}
\usepackage{CJKutf8}
\usepackage{multirow}
\usepackage{makecell}
\usepackage{pifont}
\usepackage{tikz}
\usepackage{mathrsfs}
\usepackage{threeparttable}
\usepackage{booktabs}
\usepackage{amsthm}
\usepackage{balance}

\def\BibTeX{{\rm B\kern-.05em{\sc i\kern-.025em b}\kern-.08em
    T\kern-.1667em\lower.7ex\hbox{E}\kern-.125emX}}
\makeatletter
\newenvironment{breakablealgorithm}
{
		\begin{center}
			\refstepcounter{algorithm}
			\hrule height.8pt depth0pt \kern2pt
			\renewcommand{\caption}[2][\relax]{
				{\raggedright\textbf{\ALG@name~\thealgorithm} ##2\par}%
				\ifx\relax##1\relax 
				\addcontentsline{loa}{algorithm}{\protect\numberline{\thealgorithm}##2}%
				\else 
				\addcontentsline{loa}{algorithm}{\protect\numberline{\thealgorithm}##1}%
				\fi
				\kern2pt\hrule\kern2pt
			}
		}{
		\kern2pt\hrule\relax
	\end{center}
}

\newtheorem{definition}{Definition}
\newtheorem{problem}{Problem}
\newtheorem{lemma}{Lemma}
\newtheorem{theorem}{Theorem}
\renewcommand{\IEEEproofindentspace}{0pt}

\begin{document}

\title{EVA-S2PLoR: Decentralized Secure 2-party Logistic Regression with A Subtly Hadamard Product Protocol (Full Version)}


\author{
	\IEEEauthorblockN{
		Tianle Tao\textsuperscript{1}, 
		Shizhao Peng\textsuperscript{1,2,3}, 
		Tianyu Mei\textsuperscript{1}, 
		Shoumo Li\textsuperscript{1}, 
		Haogang Zhu\textsuperscript{1,2,3,*}}\thanks{*Corresponding author.}
	\IEEEauthorblockA{\textsuperscript{1}\textit{State Key Laboratory of Complex \& Critical Software Environment, Beihang University}, Beijing, China}
	\IEEEauthorblockA{\textsuperscript{2}\textit{Zhongguancun Laboratory}, Beijing, China}
	\IEEEauthorblockA{\textsuperscript{3}\textit{Hangzhou International Innovation Institute, Beihang University}, Hangzhou, China} 
	\IEEEauthorblockA{\{taotianle, by1806167, tianyumei, 22371327, haogangzhu\}@buaa.edu.cn}
}

\maketitle

\begin{abstract}
The implementation of accurate nonlinear operators (e.g., sigmoid function) on heterogeneous datasets is a key challenge in privacy-preserving machine learning (PPML). Most existing frameworks approximate it through linear operations, which not only result in significant precision loss but also introduce substantial computational overhead. This paper proposes an efficient, verifiable, and accurate security 2-party logistic regression framework (EVA-S2PLoR), which achieves accurate nonlinear function computation through a subtly secure hadamard product protocol and its derived protocols. All protocols are based on a practical semi-honest security model, which is designed for decentralized privacy-preserving application scenarios that balance efficiency, precision, and security. High efficiency and precision are guaranteed by the asynchronous computation flow on floating point numbers and the few number of fixed communication rounds in the hadamard product protocol, where robust anomaly detection is promised by dimension transformation and Monte Carlo methods. EVA-S2PLoR outperforms many advanced frameworks in terms of precision, improving the performance of the sigmoid function by about 10 orders of magnitude compared to most frameworks. Moreover, EVA-S2PLoR delivers the best overall performance in secure logistic regression experiments with training time reduced by over 47.6\% under WAN settings and a classification accuracy difference of only about 0.5\% compared to the plaintext model.
\end{abstract}


\section{Introduction}
Classification is a core task in machine learning and remains a significant focus of current research. Logistic regression is one of the most popular classification algorithms due to its simplicity and intuitiveness. While training with extensive datasets from various domains can improve the model generalization capability, it comes with increased computational demands. To efficiently address this challenge, distributed machine learning is often employed to manage and distribute the computational workload. However, with increasing awareness of personal privacy and the introduction of privacy protection policies by governments worldwide (such as the CCPA\cite{Illman_Temple_2019} in the US), traditional training methods in fields such as finance, healthcare, and transportation introduce privacy leakage risks. For example, when using data distributed across multiple healthcare institutions to build logistic regression models for COVID-19 prevention, traditional methods typically involve uploading this distributed data to cloud servers for training. While efficient, a breach of these servers could expose patient data, causing significant financial losses. Thus, developing both secure and efficient logistic regression methods for distributed databases is of great research importance.

The emergence of privacy-preserving computation offers an effective solution to privacy protection challenges in decentralized data flow scenarios. Currently, mainstream techniques include secure multi-party computation (SMPC) \cite{Yao_1982}, homomorphic encryption (HE) \cite{Acar_Aksu_Uluagac_Conti_2018}, differential privacy (DP) \cite{Dwork_2006}, and federated learning (FL) \cite{McMahan_Moore_Ramage_Hampson_Arcas_2017}, which form the foundation of privacy-preserving machine learning (PPML).

Nonlinear operators play a crucial role in logistic regression, the hidden layers of multilayer perceptrons (MLP) \cite{Murtagh_1991}, as well as in the pooling and output layers of convolutional neural networks (CNN) \cite{Yamashita_Nishio_Do_Togashi_2018}. However, most mainstream PPML frameworks approximate nonlinear operations using linear computations (e.g., $e^x\approx 1+x+\frac{x^2}{2}+\frac{x^3}{6}$), leading to significant precision loss that affects the performance of model training and prediction. Table \ref{PPML Comparison} presents the comparison of several popular PPML frameworks that support logistic regression. 

SecureML\cite{Mohassel_Zhang_2017} was one of the earliest to propose PPML model training, including logistic regression, within a secure 2-party computation model. It approximates nonlinear activation functions using piecewise linear functions, transforming the logistic regression training problem into a combination of linear operations and comparisons. This approach leverages oblivious transfer (OT) \cite{Kilian_1988} or HE for preprocessing, secret sharing (SS) \cite{Beimel_2011} for linear operations, and garbled circuits (GC) \cite{Bellare_Hoang_Rogaway_2012} for comparisons. Although early implementations of SecureML had significant computational and communication overheads and lower accuracy due to limited optimization, its conceptual innovations have provided valuable guidance for subsequent research. ABY3\cite{Mohassel_Rindal_2018}, SecretFlow\cite{Ma_Zheng_Feng_Zhao_Wu_Fang_Tan_Yu_Zhang_Wang_2023}, BLAZE\cite{Patra_Suresh_2020}, and PrivPy\cite{Li_Xu_2019} employ optimized arithmetic and binary secret sharing, improving both computational and communication efficiency. When handling activation functions, ABY3 and BLAZE similarly use piecewise functions for approximation, while SecretFlow and PrivPy adopt polynomial approximation strategies (the Newton-Raphson method\cite{Ypma_1995} and the Euler method\cite{Stoer_Bulirsch_Bartels_Gautschi_Witzgall_1980}) to convert nonlinear functions into polynomial operations. FATE\cite{Liu_Fan_Chen_Xu_Yang_2021} utilizes FL for privacy-preserving logistic regression where data holders locally pre-train the model and send encrypted gradient information (using HE) to the arbiter for aggregation. Since HE struggles with nonlinear operations,  FATE employs the Taylor approximation\cite{Kyurkchiev_Markov_2015} to convert nonlinear functions into polynomial linear operations. MP-SPDZ\cite{Keller_2020}, based on the SPDZ-2 protocol, provides both piecewise and polynomial approximation for handling nonlinear functions, offering flexibility for users.

\begin{table*} 
  \centering
  \caption{Comparison of Various PPML Frameworks Related to Secure Logistic Regression} 
  \label{PPML Comparison}
    \resizebox{\linewidth}{!}{
    \begin{threeparttable}
    \begin{tabular}{ccccccccc}
    \toprule
    \multirow{2}[3]{*}{\textbf{Framework}} & \multicolumn{1}{c}{\multirow{2}[3]{*}{\textbf{Dev. Language}}} & \multicolumn{1}{c}{\multirow{2}[3]{*}{\textbf{Main Techniques}}} & \multicolumn{1}{c}{\multirow{2}[3]{*}{\textbf{Security Model}}} & \multicolumn{1}{c}{\multirow{2}[3]{*}{\textbf{Activation Function}*}} & \multicolumn{4}{c}{\textbf{Performance Evaluation}}  \\
\cmidrule{6-9}
    &   &   &   &   & \textbf{Comp. Complexity} & \textbf{Comm. Complexity} & \textbf{Verifiability} & \textbf{Extensibility} \\
    \midrule
    \textbf{SecureML\cite{Mohassel_Zhang_2017}} & \textbf{C++} & \textbf{SS, GC, OT, HE} & \textbf{Semi-honest} & \textbf{Piece.} & \textbf{High} & \textbf{High} & \textbf{No} & \textbf{Medium}\\
    \textbf{MP-SPDZ\cite{Keller_2020}} & \textbf{C++} & \textbf{OT, HE, SS} & \textbf{Malicious} & \textbf{Piece. or Poly.} & \textbf{High} & \textbf{High} & \textbf{No} & \textbf{Medium}\\
    \textbf{SecretFlow\cite{Ma_Zheng_Feng_Zhao_Wu_Fang_Tan_Yu_Zhang_Wang_2023}} & \textbf{C++} & \textbf{SS, GC, OT} & \textbf{Malicious} & \textbf{Poly.} & \textbf{Medium} & \textbf{Medium} & \textbf{No} & \textbf{High}\\
    \textbf{FATE\cite{Liu_Fan_Chen_Xu_Yang_2021}} & \textbf{Python} & \textbf{FL, HE, SS} & \textbf{Semi-honest} & \textbf{Poly.} & \textbf{High} & \textbf{High} & \textbf{No} & \textbf{High}\\
    \textbf{ABY3\cite{Mohassel_Rindal_2018}} & \textbf{C++} & \textbf{SS, GC, OT} & \textbf{Malicious} & \textbf{Piece.} & \textbf{Medium} & \textbf{Medium} & \textbf{No} & \textbf{Medium}\\
    \textbf{BLAZE\cite{Patra_Suresh_2020}} & \textbf{C++} & \textbf{SS, GC} & \textbf{Malicious} & \textbf{Piece.} & \textbf{Medium} & \textbf{Medium} & \textbf{No} & \textbf{Medium}\\
    \textbf{LEGO\cite{Zhou_Fu_Wei_Li_2022}} & \textbf{C++} & \textbf{SS, GC, OT, HE} & \textbf{Semi-honest} & \textbf{Piece.} & \textbf{High} & \textbf{High} & \textbf{No} & \textbf{Medium}\\
    \textbf{PrivPy\cite{Li_Xu_2019}} & \textbf{C++} & \textbf{SS, OT} & \textbf{Semi-honest} & \textbf{Poly.} & \textbf{Medium} & \textbf{Medium} & \textbf{No} & \textbf{High}\\
    \textbf{Our Work} & \textbf{Python} & \textbf{DD} & \textbf{Semi-honest} & \textbf{Exact} & \textbf{Low} & \textbf{Low} & \textbf{Yes} & \textbf{High}\\
    \bottomrule
    \end{tabular}
        \begin{tablenotes}
            \item {*We mentioned three ways of realizing the activation function, namely piecewise function approximation (\textbf{Piece.}), polynomial approximation (\textbf{Poly.}), and exact realization (\textbf{Exact}).}
        \end{tablenotes}
    \end{threeparttable}}
\end{table*}

Evidently, to achieve ideal security, most existing solutions convert calculations on the real domain $\mathbb{R}$ to computations over rings $\mathbb{Z}_{2^k}$ or finite fields $\mathbb{F}_{2^k}$, where linear approximation inevitably leads to some degree of accuracy loss and additional computational overhead. Moreover, as approximation errors accumulate, most existing frameworks lack verification algorithms to detect potential anomalies in computation. However, in most decentralized privacy-preserving applications, a model that balances efficiency, accuracy, and security is preferred over the ideal security. 

EVA-S3PC\cite{EVA-S3PC} employs data disguising (DD) technology and provides matrix-based protocols performing on floating point numbers that are efficient in computation and communication, with support for result verification. 
Leveraging the verification module and the secure 2-party matrix multiplication and hybrid multiplication (S2PM and S2PHM) protocols of EVA-S3PC, we propose the EVA-S2PLoR framework with a practical semi-honest security model, accurately implementing verifiable secure 2-party logistic regression (S2PLoR) without arbitrary linear approximations. Specifically, the main contributions of this paper are as follows:

\begin{itemize}
    \item Secure 2-party vector hadamard product (S2PHP), vector addition to product (S2PATP), vector reciprocal (S2PR), vector sigmoid (S2PS), logistic regression training (S2PLoRT) and prediction (S2PLoRP) were designed and integrated into EVA-S2PLoR, which addressed the privacy leakage issues in the result verification module of EVA-S3PC during numerical computation (e.g., hadamard product in element-wise operations).

    \item The detailed analysis of the correctness, verifiability, and complexity of the protocols were demonstrated in this paper. In addition to a formal proof based on finite fields, an insightful practical security analysis related to data distribution boundaries was also provided.
    
    \item The efficiency and precision of basic protocols were evaluated, where EVA-S2PLoR outperforms most frameworks by approximately 10 orders of magnitude in terms of precision. In addition, we compared the performance metrics of EVA-S2PLoR with three popular PPML frameworks on three common datasets, where EVA-S2PLoR achieves a classification accuracy that differs from the plaintext model by only around 0.5\%.
\end{itemize}

\textbf{Paper Organization}: The remainder of this paper is structured as follows. Section \ref{section 2: Preliminary} provides some necessary background knowledge. Section \ref{section 3: Framework} outlines the EVA-S2PLoR framework. Section \ref{section 4: Proposed Work} details the the algorithm flow of proposed protocols. Section \ref{section 5: Theoretical Analysis} presents the theoretical analysis of the proposed algorithms. Section \ref{section 6: Performance Evaluation} provides specific experiments to verify the advantages of algorithms in EVA-S2PLoR. Section \ref{section 7: Related Work} introduces the related work. Finally, Section \ref{section 8: Conclusion and Discussion} concludes the paper and discusses future work.

\section{Preliminary}\label{section 2: Preliminary}
In this section, we have done some preparatory work, including the introduction of the security model, the notation, and some algorithms and limitations of EVA-S3PC. 

\subsection{Practical Security Model}\label{security_model}
This part presents the informal definitions of the practical semi-honest security model which are inspired by Du\cite{Du_Zhan_2002}. We will continuously refine and improve this model in future.

\begin{definition}[Semi-honest adversary model \cite{Evans_Kolesnikov_Rosulek_2018}]\label{def1}
The semi-honest adversary model assumes that all participants, whether corrupted or not, follow the protocol as specified. However, participants may attempt to infer others' private data using the information they can access.
\end{definition}

\begin{definition}[Semi-honest Commodity Server]
A semi-honest Commodity Server (CS) node does not collude with any party and is only involved in the offline phase to generate data that preserves participant privacy. This node does not require any input from the participants and exits after data generation, without taking part in any further computations.
\end{definition}

\begin{definition}[Practical Security in Semi-honest Protocols]
A protocol is considered to achieve practical security if, under the semi-honest adversary model, an attacker can only reduce the possible values of a participant's private data to a range that remains acceptable within the given context, and this range either contains an infinite number of values or is sufficiently large to withstand brute-force computation.
\end{definition}

\subsection{Notation}
In this paper, all vectors are assumed to be column vectors by default (e.g., $\boldsymbol{a} \in \mathbb{R}^{n\times 1}$ is a vector of length $n$) and are allowed to be treated as single-column two-dimensional matrices for matrix operations. We use $\odot$ to denote the vector hadamard product (e.g., $\boldsymbol{a} \odot \boldsymbol{b}$), and all exponential (e.g., $e^{\boldsymbol{a}}$), logarithmic (e.g., $ln\boldsymbol{a}$), reciprocal (e.g., $\frac{1}{\boldsymbol{a}}$), and division (e.g., $\frac{\boldsymbol{a}}{\boldsymbol{b}}$) operations on vectors are performed element-wise. We use square brackets (e.g., $[c, A]$) to denote the horizontal concatenation of matrices with constants or vectors.

\subsection{Algorithms and Limitations of EVA-S3PC}
This part introduces two algorithms of EVA-S3PC\cite{EVA-S3PC} we utilize and their limitations, for which we propose corresponding solutions.
\subsubsection{Secure 2-Party Matrix Multiplication (S2PM)}
The problem definition is as follows:
\begin{problem}[S2PM]
    Alice has a matrix $A\in \mathbb{R}^{n\times s}$ and Bob has a matrix $B\in \mathbb{R}^{s\times m}$. They want to conduct the matrix multiplication, such that Alice gets $V_a\in \mathbb{R}^{n\times m}$ and Bob gets $V_b\in \mathbb{R}^{n\times m}$, where $V_a+V_b=A\times B$.
\end{problem}

\textbf{Procedure.}
The S2PM includes three stages: CS pre-processing, online computation, and result verification.

\begin{breakablealgorithm}
    \caption{S2PM CS Pre-processing Stage}
    \label{alg:S2PM-Preprocessing}
    \begin{algorithmic}[1] 
        \Require $n, s, m$
        \Ensure Alice $\Leftarrow$ $(R_a,r_a,S_t)$ and Bob $\Leftarrow$ $(R_b,r_b,S_t)$
        \State $R_a, R_b \gets$ generate random matrices
        \State $S_t = R_a \times R_b$
        \State $r_a,r_b \gets$ generate random matrices \Comment{$r_a + r_b = S_t$}
        \State \Return $(R_a, r_a, S_t)$, $(R_b, r_b, S_t)$
    \end{algorithmic}
\end{breakablealgorithm}

\begin{breakablealgorithm}
    \caption{S2PM Online Computing Stage}
    \label{alg:S2PM-Computing}
    \begin{algorithmic}[1] 
        \Require $A,R_a \in \mathbb{R}^ {n \times s}$, $B,R_b \in \mathbb{R}^{s\times m}$, $r_a,r_b\in \mathbb{R}^{n\times m}$
        \Ensure Alice $\Leftarrow(V_a,VF_a)$ and Bob $\Leftarrow(V_b,VF_b)$
        \State $\hat{A} = A + R_a$ and send $\hat{A} \Rightarrow$ Bob 
        \State $\hat{B} = B + R_b$ and send $\hat{B} \Rightarrow$ Alice 
        \State $V_b\gets$ generate a random matrix 
        \State $VF_b = V_b-\hat{A}\times B$, $T = r_b - VF_b$ 
        \State send $(VF_b,T)\Rightarrow$ Alice
        \State $V_a = T + r_a - (R_a \times \hat{B})$ 
        \State $VF_a=V_a+R_a\times \hat{B}$ and send $VF_a\Rightarrow$ Bob 
        \State \Return $(V_a,VF_a)$, $(V_b,VF_b)$ 
    \end{algorithmic}
\end{breakablealgorithm}

\textbf{Pre-processing Stage.}
The S2PM employs a semi-honest commodity server (CS) node for preprocessing. CS generates random private matrices $R_a\in \mathbb{R}^{n\times s}$ and $R_b\in \mathbb{R}^{s\times m}$, where $rank(R_a) < s$ and $rank(R_b) < s$. Subsequently, CS computes $S_t=R_a \times R_b$ and generates random matrices $r_a$, $r_b$, where $r_a + r_b = S_t$. Finally, CS sends a set of matrices $(R_a, r_a, S_t)$ to Alice and $(R_b, r_b, S_t)$ to Bob.

\textbf{Online Stage.}
Alice computes $\hat{A} = A + R_a$ and sends $\hat{A}$ to Bob while Bob computes $\hat{B} = B + R_b$ and sends $\hat{B}$ to Alice. Bob then generates a random matrix $V_b\in \mathbb{R}^ {n \times m}$, computes $VF_b=V_b-\hat{A}\times B$, $T = r_b - VF_b$, and then sends  $(VF_b,T)$ to Alice. Finally, Alice computes $V_a = T + r_a - (R_a \times \hat{B})$, $VF_a = V_a + R_a \times \hat{B}$, and sends $VF_a$ to Bob.

\textbf{Verification Stage.}
Alice and Bob perform the same steps for $l$ rounds of verification. In each round, generate a vector $\hat{\delta_a}\in \mathbb{R}^{m\times 1}$ whose elements are all randomly composed of 0 or 1, and compute $E_r=(VF_a+VF_b-S_t)\times \hat{\delta_a}$. Accept if $E_r = (0,0,\cdots,0)^T$ holds for all $l$ rounds, reject otherwise.

\begin{breakablealgorithm}
    \caption{S2PM Result Verification Stage}
    \label{alg:S2PM-Verification}
    \begin{algorithmic}[1]
        \Require $VF_a, VF_b, S_t \in \mathbb{R}^ {n \times m}$ and $l > 0$
        \Ensure Accept if verified, Reject otherwise
        \For{$i := 1$ to $l$}
            \State $\hat{\delta_a}\gets$ generate a vector randomly composed of 0 or 1 
            \State  $E_r=(VF_a+VF_b-S_t)\times \hat{\delta_a}$ 
            \If{$E_r\neq (0,0,\cdots,0)^T$}
                \State \Return Rejected; 
            \EndIf
        \EndFor
        \State \Return Accepted
    \end{algorithmic}
\end{breakablealgorithm}

\subsubsection{Secure 2-Party Matrix Hybrid Multiplication Problem (S2PHM)}
The problem definition is as follows:
\begin{problem}[S2PHM]
    Alice has private matrices $(A_1,A_2)$ and Bob has private matrices $(B_1,B_2)$, where $A_1,B_1 \in \mathbb{R}^{n\times s}$, $A_2,B_2 \in \mathbb{R}^{s\times m}$. They want to conduct the hybrid multiplication $f[(A_1, A_2), (B_1, B_2)] = (A_1 + B_1) \times (A_2+B_2)$ in which Alice gets $V_a$ and Bob gets $V_b$ such that $V_a + V_b =  (A_1 + B_1) \times (A_2+B_2)$.
\end{problem}

\textbf{Procedure.}
In S2PHM, Alice and Bob compute $V_{a0} = A_1 \times A_2, V_{b0} = B_1 \times B_2$ respectively, and then jointly compute $V_{a1} + V_{b1} = A_1 \times B_2$  and $V_{b2} + V_{a2} = B_1\times A_2$ with S2PM protocol. Finally, Alice sums $V_a = V_{a0} + V_{a1} + V_{a2}$ and Bob sums $V_b = V_{b0} + V_{b1} + V_{b2}$.

\begin{breakablealgorithm}
    \caption{S2PHM}%
    \label{alg:S2PHM}
    \begin{algorithmic}[1] 
        \Require $(A_1,B_1) \in \mathbb{R}^{n\times s}$, $(A_2,B_2) \in \mathbb{R}^{s\times m}$
        \Ensure $V_a, V_b \in\mathbb{R}^ {n \times m}$ 
        \State $V_{a0} = A_1 \times A_2$, $V_{b0} = B_1 \times B_2$ 
        \State $V_{a1}, V_{b1}\gets$ \textbf{S2PM}$(A_1, B_2)$ 
        \State $V_{b2}, V_{a2}\gets$ \textbf{S2PM}$(B_1, A_2)$ 
        \State $V_a = V_{a0} + V_{a1} + V_{a2}$, $V_b = V_{b0} + V_{b1} + V_{b2}$ 
        \State \Return $V_a,V_b$ 
    \end{algorithmic}
\end{breakablealgorithm}

\subsubsection{Limitations}
Element-wise operations on vectors or matrices play a crucial role in machine learning, with the hadamard product being a fundamental example. However, EVA-S3PC encounters significant challenges in achieving verifiable secure hadamard product computations. A seemingly intuitive approach is to replace the matrix multiplication operator in S2PM with the hadamard product operator. Nevertheless, this substitution introduces potential privacy vulnerabilities: since Alice possesses $R_a$ and $S_t$, she can infer $R_b$ from $S_t = R_a \odot R_b$ and subsequently reconstruct Bob’s private data $B$ via $\hat{B} = B + R_b$. Likewise, Bob can infer Alice’s private data $A$. Consequently, the design of EVA-S3PC is inherently constrained to matrix multiplication-based operations, and cannot achieve verifiable secure element-wise computations. This paper resolves the issue by extending the computation to a higher dimension, with details in Section \ref{section 4: Proposed Work}.

\section{Framework}\label{section 3: Framework}

\begin{figure}[ht]
  \centering
  \includegraphics[width=1.0\hsize]{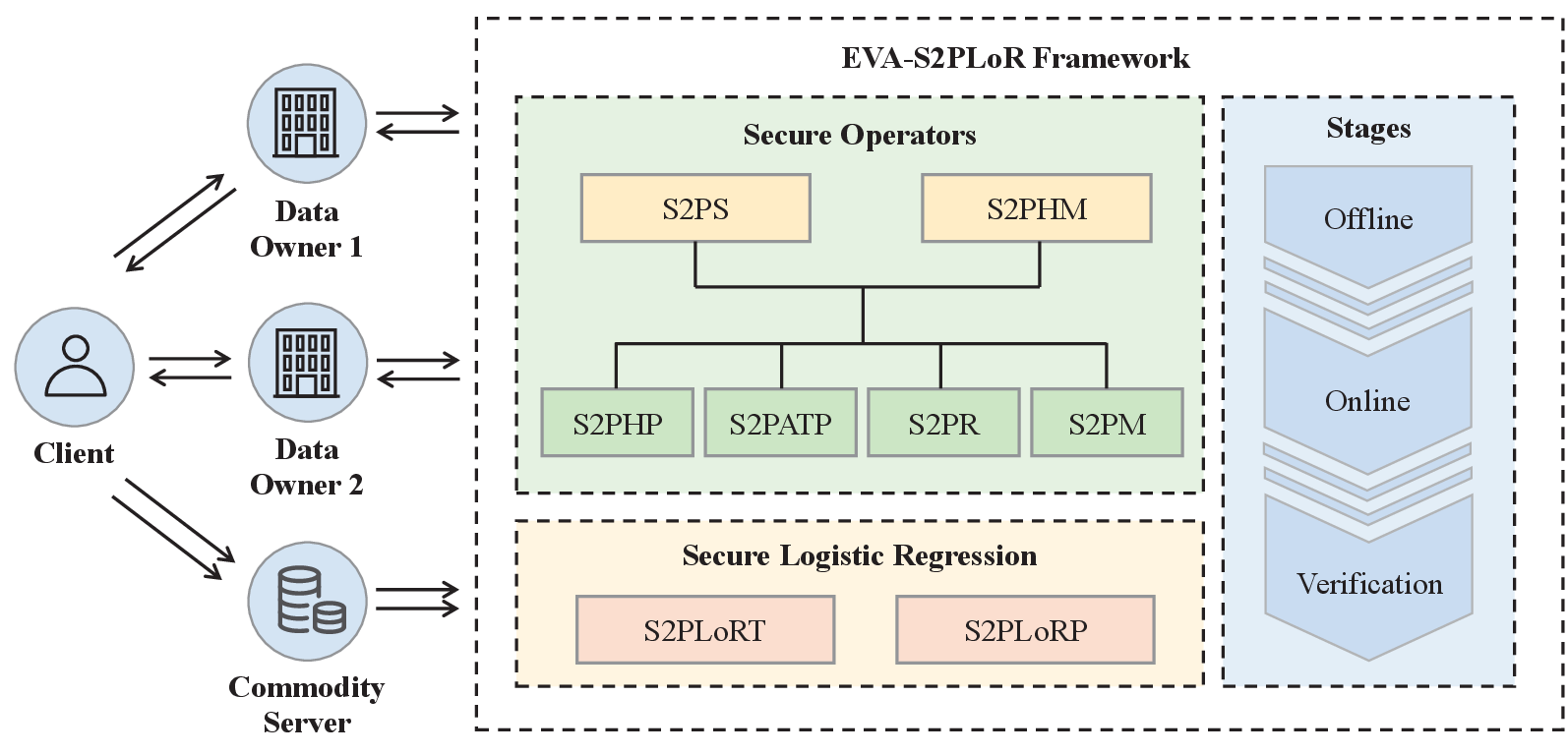}
  \caption{Framework of EVA-S2PLoR}
  \label{fig:S2PLoR-Framework}
\end{figure}

This section introduces the proposed EVA-S2PLoR framework, the structure of which is depicted in Figure \ref{fig:S2PLoR-Framework}. Intuitively, the EVA-S2PLoR framework is a decentralized computation flow composed of four real-world physical nodes, each performing different tasks. One client node initiates computation requests and retrieves the results; two data holder nodes execute the specific two-party computation protocols, and a semi-honest commodity server (CS) node performs preprocessing, neither receiving any data nor participating in online computation. Notably, the data holder nodes can also function as client nodes. In this distributed privacy computing scenario, the aforementioned nodes collaboratively implement two-party logistic regression strictly based on the secure operators (S2PHP, S2PATP, S2PR, S2PM, S2PS and S2PHM) within the EVA-S2PLoR framework. The complete computation flow can be clearly divided into three stages: offline computation, online computation, and result verification.


When the client initiates a secure logistic regression computation request, the CS node first performs preprocessing, transmits the processed data to the two data holders, and then exits the task. Subsequently, the two parties engage in data exchange and computation following the prescribed protocol. At the end of the computation, based on the data disguising mechanism, each party obtains partial model parameters, which alone do not expose the complete model. The client receives these partial results and aggregates them to construct the optimized model. In this process, the preprocessing performed by the CS node and the two data holders constitutes the offline computation phase, while the data exchange and computation between the two parties form the online computation phase. The verification phase occurs before the two parties transmit their results to the client: if the verification succeeds, the results are sent; otherwise, an anomaly is reported.


\section{Proposed Work}\label{section 4: Proposed Work}
This section presents the algorithmic procedures of the protocols in EVA-S2PLoR.

\subsection{Secure 2-Party Vector Hadamard Product (S2PHP)}

The problem definition of S2PHP is as follows:
\begin{problem}[S2PHP]\label{Problem-S2PHP}
    Alice owns a vector $\boldsymbol{a}$  and Bob owns a vector $\boldsymbol{b}$, and both vectors have length $n$. They want to conduct the vector hadamard product, such that Alice gets $\boldsymbol{v_a}$ and Bob gets $\boldsymbol{v_b}$ , where $\boldsymbol{v_a}+\boldsymbol{v_b}=\boldsymbol{a} \odot \boldsymbol{b}$.
\end{problem}

\subsubsection{Procedure}
Alice splits each $a_i \in \boldsymbol{a}$ $(1 \leq i \leq n)$ into $\rho$ random real numbers, which are uniformly sampled from a range expanded by several orders of magnitude from $a_i$'s domain (the same process is applied to Bob), forming a vector $\boldsymbol{\alpha_i}$. Then, $\rho$ identical copies of $\boldsymbol{\alpha_i}$ are concatenated to construct a vector $\boldsymbol{\alpha_i^*} \in \mathbb{R}^{\rho^2 \times 1}$. Alice then constructs a matrix $A \in \mathbb{R}^{n \times \rho^2}$ by arranging each vector $\boldsymbol{\alpha_i^*}$ sequentially as a row. Similarly, Bob splits each $b_i \in \boldsymbol{b}$ $(1 \leq i \leq n)$ into $\rho$ random real numbers, forming a vector $\boldsymbol{\beta_i}$, and then generates the full permutation of the $\rho$ elements in $\boldsymbol{\beta_i}$, denoted as $perms(\boldsymbol{\beta_i})$. From these permutations, $\rho$ vectors are randomly selected and concatenated to construct a matrix $T_i \in \mathbb{R}^{\rho \times \rho}$. Bob reshapes $T_i$ into a vector $\boldsymbol{\beta_i^*} \in \mathbb{R}^{\rho^2 \times 1}$ in a row-major order and then constructs a matrix $B \in \mathbb{R}^{\rho^2 \times n}$ by sequentially arranging each $\boldsymbol{\beta_i^*}$ as a column. Alice and Bob collaboratively compute the matrix product $A \times B = V_a + V_b$ using the S2PM protocol. They then extract the main diagonal elements of $V_a$ and $V_b$ to form vectors $\boldsymbol{v_a}$ and $\boldsymbol{v_b}$, respectively, expressed as $\boldsymbol{v_a} = diag2v(V_a)$ and $\boldsymbol{v_b} = diag2v(V_b)$. The detailed procedure is provided in Algorithm \ref{alg:S2PHP}.


\begin{breakablealgorithm}
    \caption{S2PHP}
    \label{alg:S2PHP}
    \begin{algorithmic}[1] 
        \Require $\boldsymbol{a}, \boldsymbol{b}\in\mathbb{R}^ {n\times 1}$ and  $\rho \ge 2$ 
        \Ensure $\boldsymbol{v_a} + \boldsymbol{v_b} = \boldsymbol{a} \odot \boldsymbol{b}$ where $\boldsymbol{v_a},\boldsymbol{v_b} \in\mathbb{R}^ {n\times 1} $
        \For{$i:=1$ \textbf{to} $n$}
            \State $\boldsymbol{\alpha_i} = (\alpha_{i}^{(1)}, \alpha_{i}^{(2)}, \cdots, \alpha_{i}^{(\rho)})^T$ \Comment{$\sum_{j=1}^{\rho}{\alpha_{i}^{(j)}} = a_i$}
            \State $\boldsymbol{\alpha_i^*} = [\boldsymbol{\alpha_i}^T,\boldsymbol{\alpha_i}^T,\cdots,\boldsymbol{\alpha_i}^T]^T$ 
            \State $\boldsymbol{\beta_i} = (\beta_{i}^{(1)}, \beta_{i}^{(2)}, \cdots, \beta_{i}^{(\rho)})^T$ \Comment{$\sum_{j=1}^{\rho}{\beta_{i}^{(j)}} = b_i$}
            \State $perms(\boldsymbol{\beta_i}) = \{\boldsymbol{\beta_i^{(1)}}, \boldsymbol{\beta_i^{(2)}}, \cdots, \boldsymbol{\beta_i^{(\rho!)}}\}$
            \State $T_i \gets$ concatenate $\rho$ vectors from $perms(\boldsymbol{\beta_i})$ 
            \State $\boldsymbol{\beta_i^*} \gets$ reshaped from $T_i$ in row-major order 
        \EndFor
        \State $A = [\boldsymbol{\alpha_1^*}, \boldsymbol{\alpha_2^*}, \cdots, \boldsymbol{\alpha_n^*}]^T$ 
        
        \State $B = [\boldsymbol{\beta_1^*}, \boldsymbol{\beta_2^*}, \cdots, \boldsymbol{\beta_n^*}]$
        \State $V_a, V_b\gets$ \textbf{S2PM}$(A, B)$ 
        \State $\boldsymbol{v_a} = diag2v(V_a)$, $\boldsymbol{v_b} = diag2v(V_b)$  
        \State \Return $\boldsymbol{v_a}$, $\boldsymbol{v_b}$
    \end{algorithmic}
\end{breakablealgorithm}

\subsubsection{Correctness}
We can verify that $\boldsymbol{v_a} + \boldsymbol{v_b} = diag2v(V_a+V_b)=diag2v(A\times B)=({\boldsymbol{\alpha_1^*}}^T\cdot \boldsymbol{\beta_1^*}, \cdots, {\boldsymbol{\alpha_n^*}}^T\cdot \boldsymbol{\beta_n^*})^T = (\sum_{i=1}^{\rho}{\alpha_{1}^{(i)}} \cdot \sum_{i=1}^{\rho}{\beta_{1}^{(i)}} \cdots, \sum_{i=1}^{\rho}{\alpha_{n}^{(i)}} \cdot \sum_{i=1}^{\rho}{\beta_{n}^{(i)}})^T =(a_1\cdot b_1, \cdots, a_n\cdot b_n)^T = \boldsymbol{a} \odot \boldsymbol{b}$.

\subsection{Secure 2-Party Vector Addition To Product (S2PATP)}
The problem definition of S2PATP is as follows:
\begin{problem}[S2PATP]\label{Problem-S2PATP}
    Alice owns a vector $\boldsymbol{a}$  and Bob owns a vector $\boldsymbol{b}$, and both vectors have length $n$. They want to convert the vector addition to vector hadamard product, such that Alice gets $\boldsymbol{v_a}$ and Bob gets $\boldsymbol{v_b}$ , where $\boldsymbol{v_a} \odot \boldsymbol{v_b}=\boldsymbol{a} + \boldsymbol{b}$.
\end{problem}

\subsubsection{Procedure}
Alice generates a random vector $\boldsymbol{v_a}$ of $n$ non-zero numbers and computes the vector $\boldsymbol{t_a} = \frac{1}{\boldsymbol{v_a}}$ and $\hat{\boldsymbol{a}} = \boldsymbol{a} \odot \boldsymbol{t_a}$. After that, Alice and Bob jointly compute $\boldsymbol{t_a}\odot \boldsymbol{b}=\boldsymbol{u_a} + \boldsymbol{u_b}$ with S2PHP protocol. Alice computes $\boldsymbol{t} = \hat{\boldsymbol{a}} + \boldsymbol{u_a}$ and sends $\boldsymbol{t}$ to Bob. Finally, Bob computes $\boldsymbol{v_b} = \boldsymbol{u_b} + \boldsymbol{t}$. The detailed process is shown in Algorithm \ref{alg:S2PATP}.

\begin{breakablealgorithm}
    \caption{S2PATP}
    \label{alg:S2PATP}
    \begin{algorithmic}[1] 
        \Require $\boldsymbol{a}, \boldsymbol{b}\in\mathbb{R}^ {n\times 1}$ and  $\rho \ge 2$ 
        \Ensure $\boldsymbol{v_a} \odot \boldsymbol{v_b} = \boldsymbol{a} + \boldsymbol{b}$ where $\boldsymbol{v_a},\boldsymbol{v_b} \in\mathbb{R}^ {n\times 1} $
        \State $\boldsymbol{v_a} \gets$ random vector of $n$ non-zero numbers 
        \State $\boldsymbol{t_a} = \frac{1}{\boldsymbol{v_a}}$, $\hat{\boldsymbol{a}} = \boldsymbol{a} \odot \boldsymbol{t_a}$ 
        \State $\boldsymbol{u_a}, \boldsymbol{u_b} \gets$ \textbf{S2PHP}$(\boldsymbol{t_a}, \boldsymbol{b}, \rho)$ 
        \State $\boldsymbol{t} = \hat{\boldsymbol{a}} + \boldsymbol{u_a} \Rightarrow$ Bob 
        \State $\boldsymbol{v_b} = \boldsymbol{u_b} + \boldsymbol{t}$ 
        \State \Return $\boldsymbol{v_a}$, $\boldsymbol{v_b}$
    \end{algorithmic}
\end{breakablealgorithm}

\subsubsection{Correctness}
It is easily verified that $\boldsymbol{v_a} \odot \boldsymbol{v_b} = \boldsymbol{v_a} \odot (\boldsymbol{u_b} + \boldsymbol{t}) = \boldsymbol{v_a} \odot (\boldsymbol{u_b} + \boldsymbol{u_a} + \hat{\boldsymbol{a}}) = \boldsymbol{v_a} \odot (\boldsymbol{t_a} \odot \boldsymbol{b} + \boldsymbol{t_a} \odot \boldsymbol{a}) = \boldsymbol{v_a} \odot \frac{1}{\boldsymbol{v_a}} \odot (\boldsymbol{a} + \boldsymbol{b})= \boldsymbol{a} + \boldsymbol{b}$.

\subsection{Secure 2-Party Vector Reciprocal (S2PR)}
The problem definition of S2PR is as follows:
\begin{problem}[S2PR]\label{Problem-S2PR}
    Alice owns a vector $\boldsymbol{a}$ and Bob owns a vector $\boldsymbol{b}$, and both vectors have length $n$. They want to obtain the reciprocal of the sum of two vectors, such that Alice gets $\boldsymbol{v_a}$ and Bob gets $\boldsymbol{v_b}$ , where $\boldsymbol{v_a}+\boldsymbol{v_b}=\frac{1}{\boldsymbol{a} + \boldsymbol{b}}$.
\end{problem}

\subsubsection{Procedure}
Alice and Bob jointly compute $\boldsymbol{a} + \boldsymbol{b} = \boldsymbol{u_a} \odot \boldsymbol{u_b}$ with S2PATP protocol, then Alice computes $\boldsymbol{t_a} = \frac{1}{\boldsymbol{u_a}}$ and Bob computes $\boldsymbol{t_b} = \frac{1}{\boldsymbol{u_b}}$. Finally, Alice and Bob compute $\boldsymbol{t_a} \odot \boldsymbol{t_b}=\boldsymbol{v_a} + \boldsymbol{v_b}$ with S2PHP protocol. The detailed process is shown in Algorithm \ref{alg:S2PR}.

\begin{breakablealgorithm}
    \caption{S2PR}
    \label{alg:S2PR}
    \begin{algorithmic}[1] 
        \Require $\boldsymbol{a}, \boldsymbol{b}\in\mathbb{R}^ {n\times 1}$ and  $\rho \ge 2$ 
        \Ensure $\boldsymbol{v_a} + \boldsymbol{v_b} = \frac{1}{\boldsymbol{a} + \boldsymbol{b}}$ where $\boldsymbol{v_a},\boldsymbol{v_b} \in\mathbb{R}^ {n\times 1} $
        \State $\boldsymbol{u_a}, \boldsymbol{u_b} \gets$ \textbf{S2PATP}$(\boldsymbol{a}, \boldsymbol{b}, \rho)$ 
        \State $\boldsymbol{t_a} = \frac{1}{\boldsymbol{u_a}}$, $\boldsymbol{t_b} = \frac{1}{\boldsymbol{u_b}}$ 
        \State $\boldsymbol{v_a}, \boldsymbol{v_b} \gets$ \textbf{S2PHP}$(\boldsymbol{t_a}, \boldsymbol{t_b}, \rho)$ 
        \State \Return $\boldsymbol{v_a}$, $\boldsymbol{v_b}$
    \end{algorithmic}
\end{breakablealgorithm}

\subsubsection{Correctness}
It is easily verified that $\boldsymbol{v_a} + \boldsymbol{v_b} = \boldsymbol{t_a} \odot \boldsymbol{t_b} = \frac{1}{\boldsymbol{u_a}\odot \boldsymbol{u_b}} = \frac{1}{\boldsymbol{a} + \boldsymbol{b}}$.




\subsection{Secure 2-Party Vector Sigmoid (S2PS)}
The problem definition of S2PS is as follows:
\begin{problem}[S2PS]\label{Problem-S2PS}
    Alice owns a vector $\boldsymbol{a}$  and Bob owns a vector $\boldsymbol{b}$, and both vectors have length $n$. They want to jointly conduct the sigmoid function $\sigma(x) = \frac{1}{1+e^{-x}}$. Eventually, Alice gets $\boldsymbol{v_a}$ and Bob gets $\boldsymbol{v_b}$ , where $\boldsymbol{v_a}+\boldsymbol{v_b}=\sigma (\boldsymbol{a} + \boldsymbol{b})$.
\end{problem}

\subsubsection{Procedure}
Alice and Bob respectively compute $\boldsymbol{t_a} = e^{-\boldsymbol{a}}$ and $\boldsymbol{t_b} = e^{-\boldsymbol{b}}$, and then jointly compute $\boldsymbol{t_a} \odot \boldsymbol{t_b}=\boldsymbol{u_{a}} + \boldsymbol{u_{b}}$ with S2PHP protocol. Afterwards, Alice computes $\boldsymbol{u_a^*} = \boldsymbol{u_a} + 1$. Finally, Alice and Bob jointly compute $\frac{1}{\boldsymbol{u_a^*}\ + \boldsymbol{u_b}}=\boldsymbol{v_{a}} + \boldsymbol{v_{b}}$ with S2PR protocol. The detailed process is shown in Algorithm \ref{alg:S2PS}.

\begin{breakablealgorithm}
    \caption{S2PS}
    \label{alg:S2PS}
    \begin{algorithmic}[1] 
        \Require $\boldsymbol{a}, \boldsymbol{b}\in\mathbb{R}^ {n\times 1}$ and  $\rho \ge 2$ 
        \Ensure $\boldsymbol{v_a}+\boldsymbol{v_b}=\sigma(\boldsymbol{a} + \boldsymbol{b})$ where $\boldsymbol{v_a},\boldsymbol{v_b} \in\mathbb{R}^ {n\times 1} $
        \State $\boldsymbol{t_a} = e^{-\boldsymbol{a}}$, $\boldsymbol{t_b} = e^{-\boldsymbol{b}}$ 
        \State $\boldsymbol{u_{a}}, \boldsymbol{u_{b}} \gets$ \textbf{S2PHP}$(\boldsymbol{t_a},\boldsymbol{t_b},\rho)$ 
        \State $\boldsymbol{u_a^*} = \boldsymbol{u_a} + 1$ 
        \State $\boldsymbol{v_{a}}, \boldsymbol{v_{b}} \gets$ \textbf{S2PR}$(\boldsymbol{u_a^*},\boldsymbol{u_b},\rho)$ 
        \State \Return $\boldsymbol{v_a}$, $\boldsymbol{v_b}$
    \end{algorithmic}
\end{breakablealgorithm}

\subsubsection{Correctness}
It is easily verified that $\boldsymbol{v_a} + \boldsymbol{v_b} = \frac{1}{\boldsymbol{u_a^*} + \boldsymbol{u_b}} = \frac{1}{1 + \boldsymbol{u_a} + \boldsymbol{u_b}} = \frac{1}{1 + \boldsymbol{t_a} \odot \boldsymbol{t_b}} = \frac{1}{1 + e^{-(\boldsymbol{a} + \boldsymbol{b})}} = \sigma(\boldsymbol{a} + \boldsymbol{b})$.\\




Based on the S2PS proposed above and the S2PHM in the preliminary, we can implement secure logistic regression, capable of handling heterogeneous distributed data, including but not limited to horizontal and vertical partitions. For model training, suppose that the original training dataset is $X\in\mathbb{R}^{n\times d}$, where $n$ is the number of samples and $d$ is the number of features per sample. In this paper, the heterogeneous data distribution refers to two parties that both hold a part of $X$, denoted as $X_a,X_b\in\mathbb{R}^{n\times d}$, such that $X_a + X_b = X$. The detailed design of secure logistic regression is provided below.

\subsection{Secure 2-Party Logistic Regression Training (S2PLoRT)}
The problem definition of S2PLoRT is as follows:
\begin{problem}[S2PLoRT]\label{Problem-S2PLoRT}
    For heterogeneous distributed data scenarios, Alice and Bob separately hold a partial training dataset denoted as $X_a,X_b$, and both possess the label $\boldsymbol{y}$. They aim to perform secure logistic regression training using gradient descent with the batch size of $B$, the learning rate of $\eta$, and $t$ total iterations. Finally, they individually obtain $\boldsymbol{\hat{\omega}_a}$ and $\boldsymbol{\hat{\omega}_b}$, such that $\boldsymbol{\hat{\omega}_a}+\boldsymbol{\hat{\omega}_b}=\boldsymbol{\hat{\omega}}$ where $\boldsymbol{\hat{\omega}}$ denotes the model parameter after training.
\end{problem}

\subsubsection{Procedure}
In S2PLoRT, some plaintext operations are performed first to obtain certain public variables: $N = \lceil \frac{n}{B} \rceil$ is the number of batches, $n_i(1\le i \le N)$ is the sample number in the $i$-th batch, and $\boldsymbol{y^{(i)}}$ represents the labels for the $i$-th batch. Once the above operations are completed, Alice computes $\hat{X}_a = [1, X_a]$, generates a vector $\boldsymbol{\hat{\omega}_a} = \boldsymbol{0}$ and then splits $\hat{X}_a$ into $N$ parts sequentially ($\hat{X}_a^{(i)}$ for the $i$-th batch). Meanwhile, Bob performs similar operations, except that $\hat{X}_b = [0, X_b]$.
Once the preparation is complete, the next step is to perform the gradient descent to update the parameters. Each training iteration processes all batches sequentially. For the $i$-th batch, the specific process is as follows: Alice and Bob jointly compute $ (\hat{X}_a^{(i)} + \hat{X}_b^{(i)}) \cdot (\boldsymbol{\hat{\omega}_a} + \boldsymbol{\hat{\omega}_b}) = \boldsymbol{\hat{y}_a^*} + \boldsymbol{\hat{y}_b^*}$ with S2PHM protocol, compute $\sigma(\boldsymbol{y_a^*} + \boldsymbol{y_b^*}) = \boldsymbol{\hat{y}_a} + \boldsymbol{\hat{y}_b}$ with S2PS protocol, and finally compute $\frac{1}{n_i} \cdot (\hat{X}_a^{(i)^T} + \hat{X}_b^{(i)^T}) \cdot (\boldsymbol{\hat{y}_a} - \boldsymbol{y^{(i)}} + \boldsymbol{\hat{y}_b}) = \nabla J_a +\nabla J_b$ with S2PHM protocol. Respectively, Alice and Bob update the model parameter they hold by  $\boldsymbol{\hat{\omega}}:=\boldsymbol{\hat{\omega}}-\eta \cdot \nabla J$. Training continues until $t$ iterations are completed. The detailed process is shown in Algorithm \ref{alg:S2PLoRT}.

\begin{breakablealgorithm}
    \caption{S2PLoRT}
    \label{alg:S2PLoRT}
    \begin{algorithmic}[1] 
        \Require $X_a, X_b \in \mathbb{R}^ {n\times d}, \boldsymbol{y} \in \{ 0, 1\}^ {n\times 1},t,\eta ,B,\rho$
        \Ensure $\boldsymbol{\hat{\omega}_a} + \boldsymbol{\hat{\omega}_b} = \boldsymbol{\hat{\omega}}$ where $\boldsymbol{\hat{\omega}_a}, \boldsymbol{\hat{\omega}_b}, \boldsymbol{\hat{\omega}} \in\mathbb{R}^ {(d+1)\times 1}$
        \State Preprocessing section (omitted).
        \For{$round := 1$ \textbf{to} $t$}
            \For{$i := 1$ \textbf{to} $N$}
                \State $\boldsymbol{y_a^*}, \boldsymbol{y_b^*}\gets$  \textbf{S2PHM}$(\hat{X}_a^{(i)}, \hat{X}_b^{(i)}, \boldsymbol{\hat{\omega}_a}, \boldsymbol{\hat{\omega}_b})$ 
                \State $\boldsymbol{\hat{y}_a}, \boldsymbol{\hat{y}_b}\gets$  \textbf{S2PS}$(\boldsymbol{y_a^*}, \boldsymbol{y_b^*}, \rho)$
                \State $\boldsymbol{\hat{y}_a}^*=\boldsymbol{\hat{y}_a} - \boldsymbol{y^{(i)}}$
                \State $\nabla J_a, \nabla J_b\gets \frac{1}{n_i} \cdot$  \textbf{S2PHM}$(\hat{X}_a^{(i)^T}, \hat{X}_b^{(i)^T}, \boldsymbol{\hat{y}_a}^*, \boldsymbol{\hat{y}_b})$
                \State $\boldsymbol{\hat{\omega}_a}:=\boldsymbol{\hat{\omega}_a}-\eta \cdot \nabla J_a$, $\boldsymbol{\hat{\omega}_b}:=\boldsymbol{\hat{\omega}_b}-\eta \cdot \nabla J_b$ 
            \EndFor
        \EndFor
        \State \Return $\boldsymbol{\hat{\omega}_a}$, $\boldsymbol{\hat{\omega}_b}$
    \end{algorithmic}
\end{breakablealgorithm}

\subsubsection{Correctness}
For convenience, assume that each round of training uses the full training set (i.e., $B = n$). For each round, let $\boldsymbol{\hat{\omega}_a} + \boldsymbol{\hat{\omega}_b}=\boldsymbol{\hat{\omega}}$, then it is easily verified that $\boldsymbol{\hat{y}_a} + \boldsymbol{\hat{y}_b} = \sigma(\boldsymbol{y_a^*} + \boldsymbol{y_b^*}) = \sigma((\hat{X}_a + \hat{X}_b)\cdot (\boldsymbol{\hat{\omega}_a} + \boldsymbol{\hat{\omega}_b})) = \sigma(\hat{X}\cdot \boldsymbol{\hat{\omega}}) = \boldsymbol{\hat{y}}$, and thus $\boldsymbol{\hat{\omega}_a} + \boldsymbol{\hat{\omega}_b}:=\boldsymbol{\hat{\omega}_a} + \boldsymbol{\hat{\omega}_b} - \eta \cdot (\nabla J_a+\nabla J_b) = \boldsymbol{\hat{\omega}_a} + \boldsymbol{\hat{\omega}_b} - \eta \cdot \frac{1}{n_i} \cdot((\hat{X}_a^{^T} + \hat{X}_b^{^T})\cdot (\boldsymbol{\hat{y}_a} + \boldsymbol{\hat{y}_b} - \boldsymbol{y})) = \boldsymbol{\hat{\omega}_a} + \boldsymbol{\hat{\omega}_b} - \eta \cdot \frac{1}{n_i} \cdot \hat{X}^{^T} \cdot (\boldsymbol{\hat{y}}- \boldsymbol{y})$. Finally, it can be seen that the formula for the gradient update $\boldsymbol{\hat{\omega}} := \boldsymbol{\hat{\omega}} - \eta \cdot \frac{1}{n_i} \cdot \hat{X}^{^T} \cdot (\boldsymbol{\hat{y}}- \boldsymbol{y})$ holds.




\subsection{Secure 2-Party Logistic Regression Prediction (S2PLoRP)}
The problem definition of S2PLoRP is as follows:
\begin{problem}[S2PLoRP]\label{Problem-S2PLoRP}
    For heterogeneous distributed data scenarios, Alice and Bob separately hold a portion of the predicting dataset denoted as $X_a^*,X_b^*$, and the model parameters denoted as $\boldsymbol{\hat{\omega}_a}, \boldsymbol{\hat{\omega}_b}$. They aim to perform the privacy-preserving logistic regression prediction such that Alice gets $\boldsymbol{\hat{y}_a}$ and Bob gets $\boldsymbol{\hat{y}_b}$, where $\boldsymbol{\hat{y}_a} + \boldsymbol{\hat{y}_b} = \boldsymbol{\hat{y}}$ and $\boldsymbol{\hat{y}}$ denotes the result of model prediction.
\end{problem}

\subsubsection{Procedure}
In S2PLoRP, similar to S2PLoRT, Alice and Bob compute $\hat{X}_a^* = [1, X_a^*] ,\hat{X}_b^* = [0, X_b^*]$ respectively and then jointly compute $ (\hat{X}_a^{*} + \hat{X}_b^{*}) \cdot (\boldsymbol{\hat{\omega}_a} + \boldsymbol{\hat{\omega}_b}) = \boldsymbol{\hat{y}_a^*} + \boldsymbol{\hat{y}_b^*}$ with S2PHM protocol, and finally compute $\sigma(\boldsymbol{y_a^*} + \boldsymbol{y_b^*}) = \boldsymbol{\hat{y}_a} + \boldsymbol{\hat{y}_b}$ with S2PS protocol. The detailed process is shown in Algorithm \ref{alg:S2PLoRP}.

\begin{breakablealgorithm}
    \caption{S2PLoRP}
    \label{alg:S2PLoRP}
    \begin{algorithmic}[1] 
        \Require $X_a^*, X_b^* \in \mathbb{R}^ {n\times d}, \boldsymbol{\hat{\omega}_a}, \boldsymbol{\hat{\omega}_b} \in \mathbb{R}^{(d+1) \times 1}$, and $\rho \ge 2$.
        \Ensure $\boldsymbol{\hat{y}_a} + \boldsymbol{\hat{y}_b} = \boldsymbol{\hat{y}}$ where $\boldsymbol{\hat{y}_a}, \boldsymbol{\hat{y}_b}, \boldsymbol{\hat{y}} \in \mathbb{R}^{n\times 1}$
        \State $\hat{X}_a^* = [1, X_a^*]$ and $\hat{X}_b^* = [0, X_b^*]$ 
        \State $\boldsymbol{y_a^*}, \boldsymbol{y_b^*}\gets$  \textbf{S2PHM}$(\hat{X}_a^{*}, \hat{X}_b^{*}, \boldsymbol{\hat{\omega}_a}, \boldsymbol{\hat{\omega}_b})$
        \State $\boldsymbol{\hat{y}_a}, \boldsymbol{\hat{y}_b}\gets$  \textbf{S2PS}$(\boldsymbol{y_a^*}, \boldsymbol{y_b^*}, \rho)$ 
        \State \Return $\boldsymbol{\hat{y}_a}$, $\boldsymbol{\hat{y}_b}$
    \end{algorithmic}
\end{breakablealgorithm}

\subsubsection{Correctness}
It is easily verified that $\boldsymbol{\hat{y}_a} + \boldsymbol{\hat{y}_b} = \sigma(\boldsymbol{y_a^*} + \boldsymbol{y_b^*}) = \sigma((\hat{X}_a^* + \hat{X}_b^*)\cdot (\boldsymbol{\hat{\omega}_a} + \boldsymbol{\hat{\omega}_b})) = \sigma(\hat{X}^* \cdot \boldsymbol{\hat{\omega}}) = \boldsymbol{\hat{y}}$.




\section{Theoretical Analysis}\label{section 5: Theoretical Analysis}
In this section, we provide a theoretical analysis on the verifiability, security, computational and communication complexity of proposed protocols.

\subsection{Verifiability Analysis}
The verification module of EVA-S2PLoR is provided by the underlying S2PM protocol (verification algorithm detailed in Preliminary) and is primarily designed to detect rare computational anomalies during protocol execution. Additionally, it helps mitigate deviations from the prescribed protocol flow by participants, ensuring the correctness of the protocol. In EVA-S3PC, it has been proven that the probability of S2PM failing to detect an anomaly is $P_f(S2PM)\le 4^{-l}$, where $l$ represents the number of verification rounds (see Appendix \ref{append: verification proof} for proof). In practical applications, setting $l=20$ balances verification overhead and success rate, resulting in $P_f(S2PM) \le 4^{-20} \approx 9.09 \times 10^{-13}$, which is negligibly small.

Due to the independence of the verification module, a protocol fails to detect anomalies if and only if all its sub-protocols fail to do so. Therefore, the failure probabilities of each protocol in EVA-S2PLoR are displayed in Table \ref{tab:Verifiability Analysis}.

\begin{table}[htbp]
  \centering
  \caption{Analysis of Verification Failure Probability}
    \resizebox{\linewidth}{!}{
    \begin{tabular}{cccc}
    \toprule
    \multirow{2}[2]{*}{\textbf{Protocol}} & 
    \multirow{2}[2]{*}{\textbf{Probability Formula}} &
    \multicolumn{2}{c}{\textbf{Failure Probability} ($\le$)}\\
    \cmidrule{3-4}
    & & \textbf{Theoretical} & \textbf{Practical} ($l=20$) \\
    \midrule
    \textbf{S2PM} & / & $4^{-l}$ & $9.09\times10^{-13}$ \\
    \textbf{S2PHM} & $P_f(S2PM)^2$ & $4^{-2l}$ & $8.27 \times 10^{-25}$ \\
    \textbf{S2PHP} & $P_f(S2PM)$ & $4^{-l}$ & $9.09\times10^{-13}$ \\
    \textbf{S2PATP} & $P_f(S2PHP)$ & $4^{-l}$ & $9.09\times10^{-13}$ \\
    \textbf{S2PR} & $P_f(S2PATP)\cdot P_f(S2PHP)$ & $4^{-2l}$ & $8.27 \times 10^{-25}$ \\
    \textbf{S2PS} & $P_f(S2PHP)\cdot P_f(S2PR)$ & $4^{-3l}$ & $7.52 \times 10^{-37}$ \\
    \textbf{S2PLoRT} & $P_f(S2PHM)^2\cdot P_f(S2PS)$ & $4^{-7l}$ & $5.15 \times 10^{-85}$ \\
    \textbf{S2PLoRP} & $P_f(S2PHM)\cdot P_f(S2PS)$ & $4^{-5l}$ & $6.22 \times 10^{-61}$ \\
    \bottomrule
    \end{tabular}}%
  \label{tab:Verifiability Analysis}%
\end{table}%

\subsection{Security Analysis}
\subsubsection{Practical Security Analysis Based on Floating-point Numbers}
First, analyze the privacy leakage scenarios in the fundamental protocol S2PHP. In S2PHP, Alice's private data consists of the input 
$\boldsymbol{a}$ and the output $\boldsymbol{v_a}$. The set of messages received by Alice is given by $M_A=\{R_a, r_a, S_t, \hat{B}, T, VF_b\}$, and the intermediate values generated during computation are $I_A = \{A, \hat{A}, V_a, VF_a\}$. Similarly, Bob's private data consists of the input $\boldsymbol{b}$ and the output $\boldsymbol{v_b}$. The set of messages received by Bob is $M_B=\{R_b, r_b, S_t, \hat{A}, VF_a\}$, and the intermediate values generated during computation are $I_B=\{B, \hat{B}, T, V_b, VF_b\}$. Under the security constraint that both $R_a$ and $R_b$ are rank-deficient matrices, it follows that the set of data Alice cannot infer is $U_A=\{\boldsymbol{b}, B, R_b, V_b, \boldsymbol{v_b}\}$, while the set of data Bob cannot infer is $U_B=\{\boldsymbol{a}, A, R_a, V_a, \boldsymbol{v_a}\}$. A detailed analysis of privacy leakage is provided in Table \ref{tab:Privacy Leakage Analysis}.

\begin{table}[htbp]
  \centering
  \caption{Privacy Leakage Analysis of S2PHP}
    \resizebox{\linewidth}{!}{
    \begin{tabular}{cccc}
    \toprule
    \textbf{Participant} & 
    \textbf{Private Data} &
    \textbf{Indeterminable Data} &
    \textbf{Security Constraints} \\
    \midrule
    \textbf{Alice} & $\{\boldsymbol{a},\boldsymbol{v_a}\}$ & $\{\boldsymbol{b}, B, R_b, V_b, \boldsymbol{v_b}\}$ & $R_a$ is rank-deficient \\
    \textbf{Bob} & $\{\boldsymbol{b},\boldsymbol{v_b}\}$ & $\{\boldsymbol{a}, A, R_a, V_a, \boldsymbol{v_a}\}$ & $R_b$ is rank-deficient \\
    \bottomrule
    \end{tabular}}%
  \label{tab:Privacy Leakage Analysis}%
\end{table}%

While analysis shows that S2PHP does not expose participants' private data under security constraints, some privacy information inevitably leaks due to the protocol's reliance on floating-point arithmetic. As an example, consider Bob attempting to infer Alice's private data $\boldsymbol{a}$ (which becomes $A$ after preprocessing). Two equations could potentially lead to data leakage: $\hat{A}=A+R_a$ and $R_a\times R_b = S_t$. When Bob receives the message $\hat{A}$ sent by Alice, he can infer information about Alice's original data $A$ based on the equation $\hat{A}=A+R_a$. For simplicity, consider the matrices here as scalar values. If Bob knows the range of $A$ to be $[l_1, r_1](r_1\ge l_1)$, and the range of $R_a$ to be $[l_2, r_2](r_2\ge l_2)$, Bob can use the value of $\hat{A}$ to narrow down the possible range of $A$ to $[l_1, r_1]\cap [\hat{A} - r_2, \hat{A} - l_2]$. Clearly, Bob cannot shrink the range of $A$ when $\hat{A} - r_2 \le l_1$ and $\hat{A} - l_2 \ge r_1$ (i.e., when $l_2 + r_1 \le \hat{A}\le l_1+r_2$). Let $L_1 = r_1-l_1$, $L_2 = r_2-l_2$, $\theta = \frac{L2}{L1}$. When $A$ is fixed, if $R_a$ is uniformly sampled from the range $[l_2,r_2]$, then $\hat{A}$ is uniformly distributed over $[l_1+l_2, r_1+r_2]$. The probability that Bob cannot narrow down the range of $A$:
\begin{align}
P=\frac{(l_1+r_2) - (l_2+r_1)}{(r_1+r_2) - (l_1+l_2)}=\frac{(L_2-L_1)}{(L_2+L_1)}=1 - \frac{2}{\theta +1}
\end{align}
When $\theta\ge 10^{4}$, $P\ge99.98\%$, which is sufficiently large to almost guarantee that Bob will not gain any additional information about $A$. Bob can also infer the range of $R_a$ as $[l_2, r_2]\cap [\hat{A} - r_1, \hat{A} - l_1]$, and the probability of narrowing the range of $R_a$ to at least $L_1$ (i.e., $\hat{A} - r_1\ge l_2$ and $\hat{A} - l_1 \le r_2$) is similarly $P= 1 - \frac{2}{\theta +1}$. At this point, the range of $R_a$ is at least as large as the range of the private data $A$, ensuring practical security. When $R_a$ and $R_b$ are rank-deficient matrices, $R_a\times R_b = S_t$ forms an underdetermined matrix equation. As shown in \cite{ben2006generalized}, when Bob holds $R_b$ and $S_t$, he can deduce $R_a = St\times R_b^+ + N$, where $R_b^+$ is the Moore-Penrose pseudoinverse of $R_b$, and $N=Z\times C$ represents the left null space of $R_b$. Since $C$ is a free variable, $R_a$ has infinite solutions, and the range of possible values is sufficiently large. In summary, we can adjust the value of $\theta$ to ensure that the S2PHP satisfies practical security. Similarly, we can prove that all other protocols in EVA-S2PLoR also meet practical security requirements, leading to the following Theorem \ref{theorem: Protocol Security}.

\begin{theorem}\label{theorem: Protocol Security}
    The protocols in EVA-S2PLoR framework is secure in the practical semi-honest security model.
\end{theorem}

\subsubsection{Indistinguishability Security Analysis Based on Finite Fields}
We also use indistinguishability to provide the proof of the aforementioned Theorem \ref{theorem: Protocol Security}.

\begin{IEEEproof}
See Appendix \ref{append: security proof}.
\end{IEEEproof}

\subsection{Complexity Analysis}
For convenience, assume that all the vectors in basic protocols are of length $n$ with the split factor $\rho$, and each element is encoded in length $\ell$. Moreover, assume that all the datasets in logistic regression are matrices of shape $n\times d$ and the model is trained with a complete dataset (i.e., $B=n$) in each of $t$ total iterations, without a bias term. The detailed results of the analysis are displayed in Table \ref{tab:Complexity Analysis}. 

\begin{table}[htbp]
  \centering
  \caption{Complexity Analysis of EVA-S2PLoR}
    \resizebox{\linewidth}{!}{
    \begin{tabular}{cccc}
    \toprule
    \textbf{Protocol} & 
    \textbf{Computational Complexity} &
    \textbf{Communication Cost [bits]} &
    \textbf{Communication Rounds} \\
    \midrule
    \textbf{S2PHP} & $\mathcal{O}(n^2\rho^2)$ &  $(4n\rho^2+7n^2)\ell$ &   6 \\
    
    \textbf{S2PATP} & $\mathcal{O}(n^2\rho^2)$ &  $(4n\rho^2+7n^2 + n)\ell$ &   7 \\

    \textbf{S2PR} & $\mathcal{O}(n^2\rho^2)$ &  $(8n\rho^2+14n^2 + n)\ell$ &   13 \\

   \textbf{S2PS} & $\mathcal{O}(n^2\rho^2)$ &   $(12n\rho^2+21n^2 + n)\ell$ &   19 \\

    \textbf{S2PLoRT} & $\mathcal{O}(n^2\rho^2t + ndt)$ & $(8nd+18d+12n\rho^2+21n^2 + 19n)t\ell$ &    43$t$ \\

    \textbf{S2PLoRP} & $\mathcal{O}(n^2\rho^2 + nd)$ &     $(4nd+4d+12n\rho^2+21n^2 + 15n)\ell$ &  31 \\
    \bottomrule
    \end{tabular}}%
  \label{tab:Complexity Analysis}%
\end{table}%

Although these protocols for security vector operations appear to be of fourth-order complexity, in practice, the values of $\rho$ are usually small and can be considered as constants. We can optimize by batching long vectors and performing parallel computations to reduce the computational and communication overhead caused by dimensional transformations.


\section{Performance Evaluation}\label{section 6: Performance Evaluation}
In this section, we evaluate the performance of proposed protocols in the EVA-S2PLoR framework.

\subsection{Settings}
All the protocols proposed in this paper are implemented in Python, and each protocol can be viewed as a separate module. Performance evaluation experiments for the EVA-S2PLoR framework were conducted on a machine with 22 vCPUs (Intel$^\circledR$ Core\texttrademark\ Ultra 7), 32 GB RAM, and Ubuntu 22.04 LTS.
In order to minimize the inconsistency caused by different communication environments, we set the network bandwidth to 10.1/300 Gbps and the latency to 0.1/40 ms for the LAN/WAN environment. In the performance evaluation experiment of the EVA-S2PLoR framework, four nodes, as mentioned in Section \ref{section 3: Framework}, are simulated and started via ports.

\subsection{Performance of Basic Protocols}
We conducted experiments for the computational and communication overhead, and precision loss of basic protocols in EVA-S2PLoR. Since most existing frameworks implement secure logistic regression using approximation methods, which differ from approaches in EVA-S2PLoR, they do not have or need not implement S2PATP and S2PR protocols, and therefore we focused on comparing the S2PHP and S2PS protocols. For S2PHP, we compared four representative privacy-preserving computing solutions in a semi-honest environment: CrypTen (SS), ABY (GC), LibOTe (OT), and TenSEAL (HE). For S2PS, we additionally included a comparison with the MP-SPDZ framework, which combines multiple technologies.

Considering that most existing frameworks perform computations in the ring domain $\mathbb{Z}_{2^k}$ as well as the limitations of computers in storing and representing floating-point numbers, we use vectors randomly composed of elements with 16 significant digits that can be represented as $\pm1.a_1a_2\cdots a_{15}\times 10^\delta (\delta \in [-x,x], x\in \mathbb{Z})$ as input. In efficiency evaluation, we set $x = 4$, and in precision evaluation, we let $x$ increase in steps of 2 from 0 to 8 thus dividing $\delta$ into five ranges.

\subsubsection{Efficiency Evaluation}
\begin{figure}[htb]
    \centering
    
    \subfigure[S2PHP Comm. Overhead]
    {\includegraphics[width=0.45\linewidth]{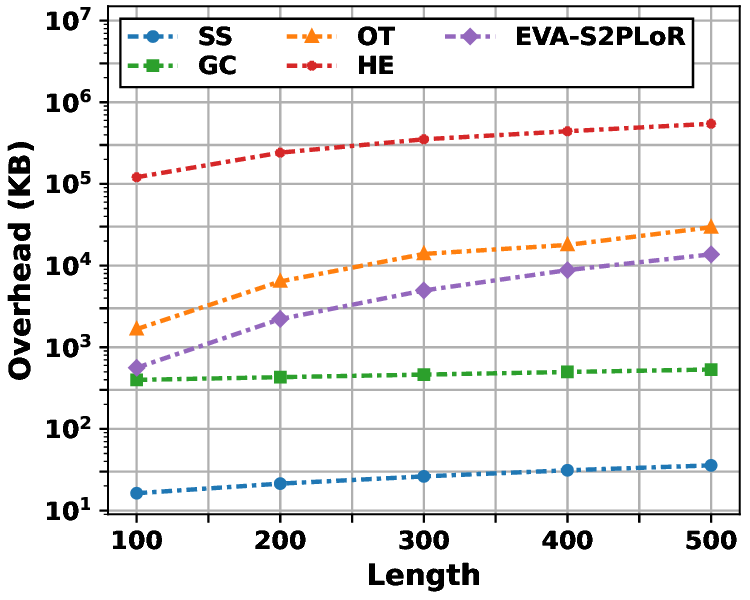}
    \label{fig: S2PHP Comm.Overhead}}
    \hspace{0.3cm}
    \subfigure[S2PS Comm. Overhead]{\includegraphics[width=0.45\linewidth]{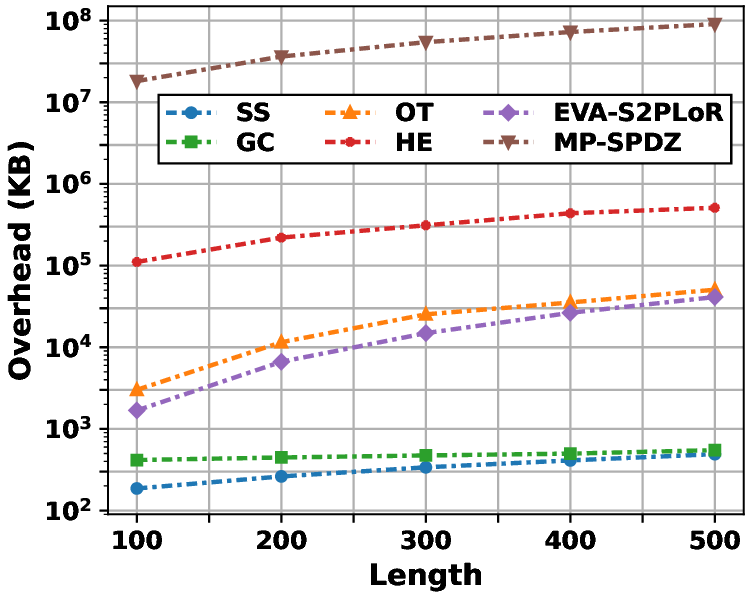}
    \label{fig: S2PS Comm.Overhead}}
    

    \centering
    \subfigure[S2PHP Running Time (LAN)]{\includegraphics[width=0.45\linewidth]{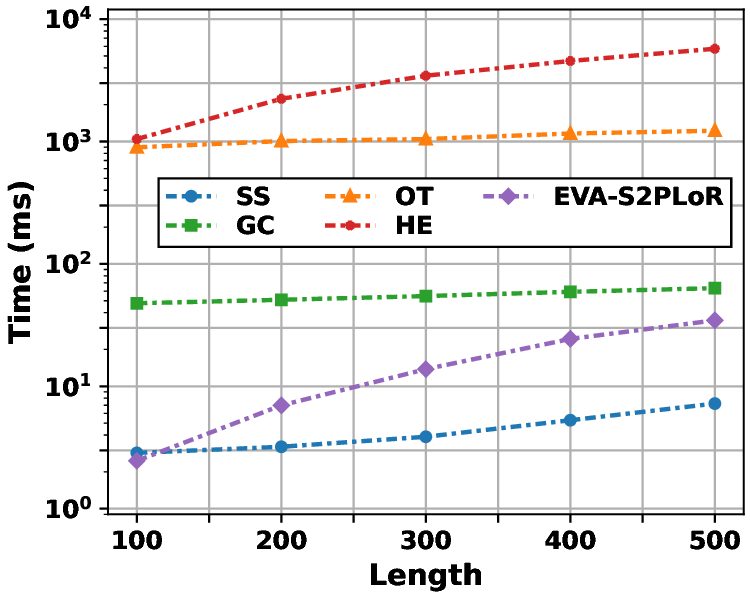}
    \label{fig: S2PHP Running Time (LAN)}}
    \hspace{0.3cm}
    \subfigure[S2PS Running Time (LAN)]{\includegraphics[width=0.45\linewidth]{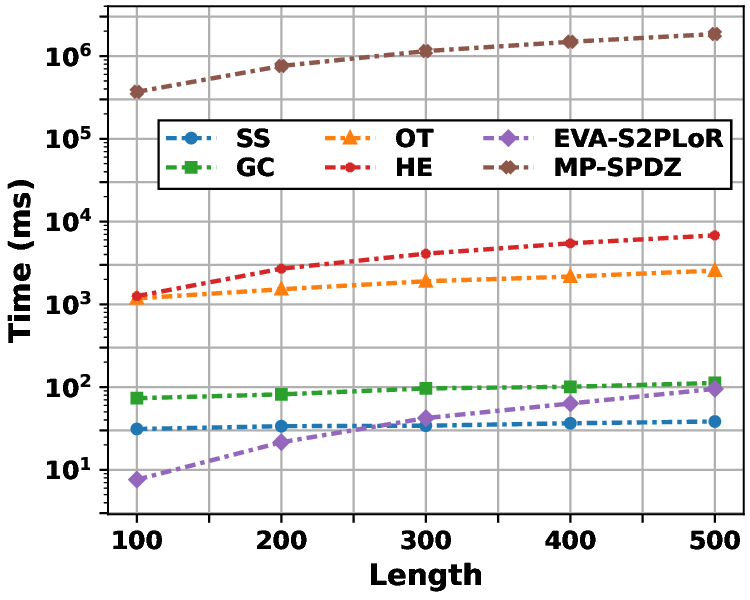}
    \label{fig: S2PS Running Time (LAN)}}

     \centering
    \subfigure[S2PHP Running Time (WAN)]{\includegraphics[width=0.45\linewidth]{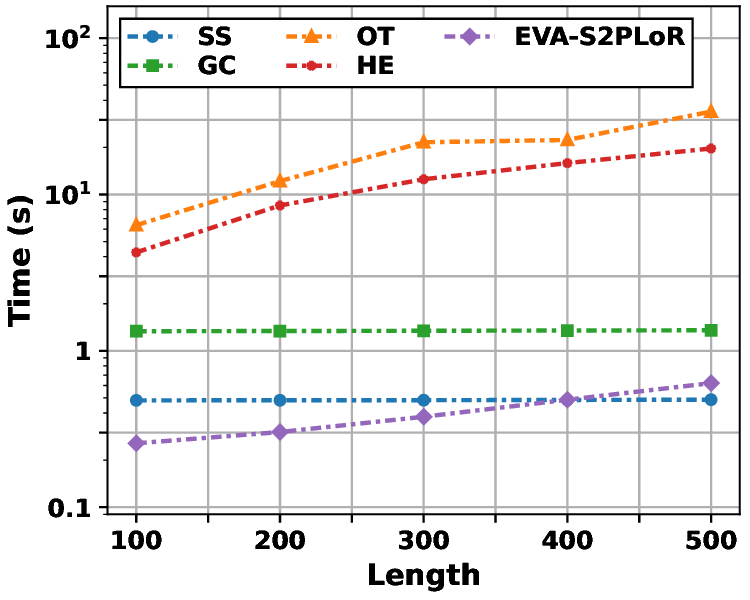}
    \label{fig: S2PHP Running Time (WAN)}}
    \hspace{0.3cm}
    \subfigure[S2PS Running Time (WAN)]{\includegraphics[width=0.45\linewidth]{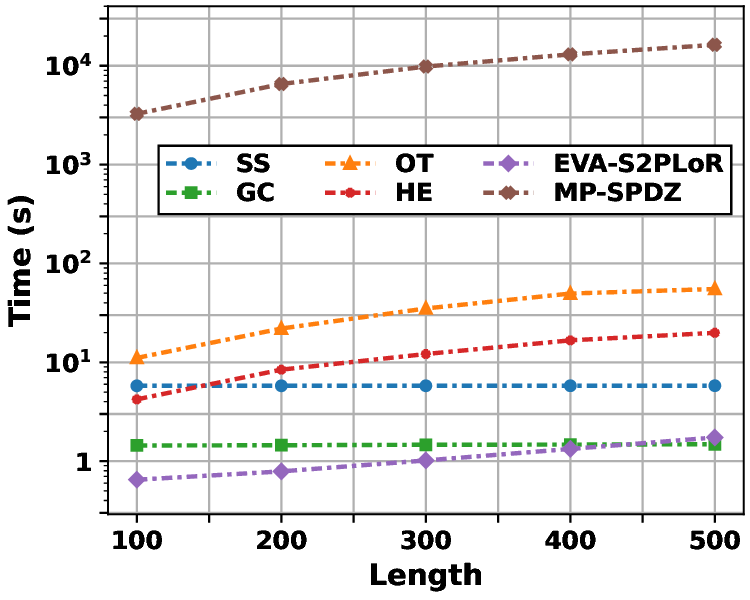}
    \label{fig: S2PS Running Time (WAN)}}
    
    \caption{Efficiency comparison in S2PHP and S2PS with SS, GC, OT, HE and MP-SPDZ.}
    \label{fig: Efficiency Comparison in S2PHP and S2PS}
\end{figure}

\begin{figure}[htb]
    \centering
    \subfigure[Proportion in S2PS]{\includegraphics[width=0.45\linewidth]{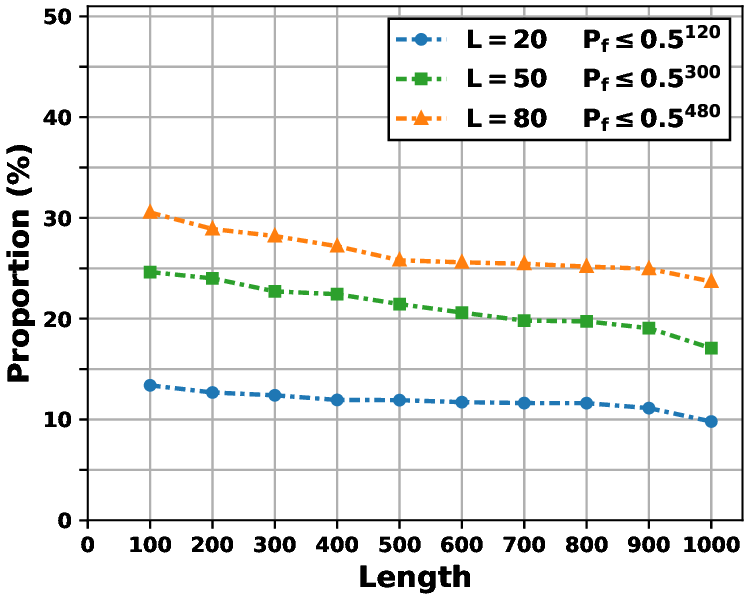}
    \label{fig: Proportion in S2PS}}
    \hspace{0.3cm}
    \subfigure[Proportion in S2PHM]{\includegraphics[width=0.45\linewidth]{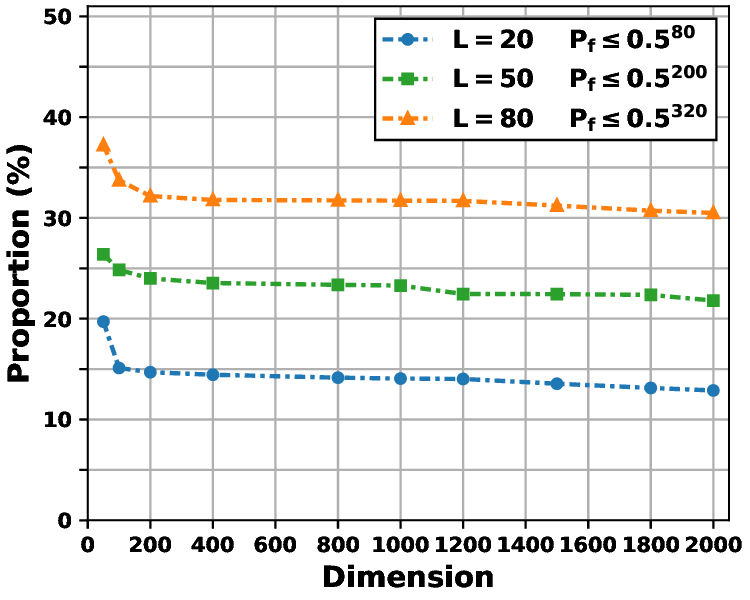}
    \label{fig: Proportion in S2PHM}}
    \caption{The impact of verification rounds $L$ on the proportion of verification time across different data dimensions in S2PS and S2PHM (key protocols to implement S2PLoR).}
    \label{fig: Proportion of Verification Cost in S2PS and S2PHM}
    \vspace{-0.2cm}
\end{figure}

We first evaluated the efficiency of basic protocols in EVA-S2PLoR and the average results are displayed in Table \ref{tab:Our-Time}. Subsequently, we compared with mainstream frameworks on S2PHP and S2PS in Fig. \ref{fig: Efficiency Comparison in S2PHP and S2PS} and evaluated the verification impact in Fig. \ref{fig: Proportion of Verification Cost in S2PS and S2PHM}.

\begin{table}[htbp]
  \centering
  \caption{Efficiency of basic protocols under LAN and WAN where 2000 repetitive experiments are conducted in terms of computation time and communication overhead.}
  \resizebox{\linewidth}{!}{
    \begin{tabular}{ccccccccccccc}
    \toprule  
    \multicolumn{1}{c}{\multirow{2}[2]{*}{\textbf{Protocol}}} & 
    \multicolumn{1}{c}{\multirow{2}[2]{*}{\textbf{Length}}} & 
    \multicolumn{2}{c}{\textbf{Communication}} & & 
    \multicolumn{4}{c}{\textbf{Computation (s)}} & & 
    \multicolumn{2}{c}{\textbf{Running Time (s)}} \\
    \cmidrule{3-4} \cmidrule{6-9} \cmidrule{11-12}
    &  & {\textbf{Comm. (KB)}} & {\textbf{Rounds}} & & \textbf{Offline} & 
    \textbf{Online} & \textbf{Verification} & \textbf{Total} & & {\textbf{LAN}} & {\textbf{WAN}} \\
    \midrule
    \multirow{4.5}[2]{*}{\textbf{S2PHP}} & 
    100 & {560.52} & {\multirow{5}[2]{*}{6}} & & 9.67E-04 & 4.09E-04 & 6.18E-05 & 1.44E-03 & & {2.46E-03} & {2.56E-01} \\
    & 200 & {2213.64} & & & 2.62E-03 & 1.95E-03 & 1.43E-04 & 4.72E-03 & & {6.99E-03} & {3.02E-01} \\
    & 300 & {4960.53} & & & 4.84E-03 & 4.37E-03 & 2.44E-04 & 9.46E-03 & & {1.38E-02} & {3.79E-01} \\
    & 400 & {8801.16} & & & 8.57E-03 & 8.10E-03 & 4.75E-04 & 1.71E-02 & & {2.44E-02} & {4.86E-01} \\
    & 500 & {13735.53} & & & 1.05E-02 & 1.23E-02 & 7.61E-04 & 2.36E-02 & & {3.46E-02} & {6.21E-01} \\
    \midrule
    \multirow{4.5}[2]{*}{\textbf{S2PATP}} & 
    100 & {561.45} & {\multirow{5}[2]{*}{7}} & & 1.34E-03 & 4.66E-04 & 6.41E-05 & 1.87E-03 & & {2.99E-03} & {2.96E-01} \\
    & 200 & {2215.35} & & & 3.66E-03 & 2.57E-03 & 2.16E-04 & 6.44E-03 & & {8.82E-03} & {3.44E-01} \\
    & 300 & {4963.03} & & & 6.35E-03 & 5.50E-03 & 2.51E-04 & 1.21E-02 & & {1.66E-02} & {4.21E-01} \\
    & 400 & {8804.43} & & & 8.45E-03 & 9.94E-03 & 5.59E-04 & 1.90E-02 & & {2.63E-02} & {5.28E-01} \\
    & 500 & {13739.59} & & & 1.30E-02 & 1.31E-02 & 8.39E-04 & 2.70E-02 & & {3.80E-02} & {6.65E-01} \\
    \midrule
    \multirow{4.5}[2]{*}{\textbf{S2PR}} & 
    100 & {1121.72} & {\multirow{5}[2]{*}{11}} & & 2.32E-03 & 9.46E-04 & 1.15E-04 & 3.38E-03 & & {5.33E-03} & {4.73E-01} \\
    & 200 & {4428.75} & & & 5.82E-03 & 4.71E-03 & 2.73E-04 & 1.08E-02 & & {1.52E-02} & {5.66E-01} \\
    & 300 & {9923.32} & & & 1.12E-02 & 1.06E-02 & 5.23E-04 & 2.24E-02 & & {3.10E-02} & {7.21E-01} \\
    & 400 & {17605.35} & & & 1.50E-02 & 1.56E-02 & 1.06E-03 & 3.16E-02 & & {4.60E-02} & {9.30E-01} \\
    & 500 & {27474.88} & & & 1.96E-02 & 2.29E-02 & 1.44E-03 & 4.40E-02 & & {6.58E-02} & {1.20E+00} \\
    \midrule
    \multirow{4.5}[2]{*}{\textbf{S2PS}} & 
    100 & {1682.00} & {\multirow{5}[2]{*}{15}} & & 3.23E-03 & 1.46E-03 & 1.51E-04 & 4.84E-03 & & {7.61E-03} & {6.49E-01} \\
    & 200 & {6642.16} & & & 8.20E-03 & 6.42E-03 & 3.99E-04 & 1.50E-02 & & {2.15E-02} & {7.88E-01} \\
    & 300 & {14883.62} & & & 1.54E-02 & 1.33E-02 & 7.45E-04 & 2.94E-02 & & {4.22E-02} & {1.02E+00} \\
    & 400 & {26406.27} & & & 2.02E-02 & 2.02E-02 & 1.45E-03 & 4.19E-02 & & {6.33E-02} & {1.33E+00} \\
    & 500 & {41210.18} & & & 3.01E-02 & 3.01E-02 & 2.33E-03 & 6.25E-02 & & {9.51E-02} & {1.74E+00} \\
    \bottomrule
    \end{tabular}}%
  \label{tab:Our-Time}%
\end{table}%

Table \ref{tab:Our-Time} illustrates that even with the introduction of the verification mechanism, these basic protocols still run fast in the LAN and WAN by virtue of their simple computational flow. Note that the communication rounds for the S2PR and S2PS protocols are less than that shown in the theoretical analysis in Table \ref{tab:Complexity Analysis}, because we send all the random matrices generated in the S2PM CS-Preprocessing Stage to the two participants through one communication, respectively, thus reducing the number of communication rounds.

As shown in Fig. \ref{fig: Efficiency Comparison in S2PHP and S2PS}, SS achieves the optimal performance on S2PHP and S2PS, due to the fact that it requires fewer secret sharing operations and less overall communication overhead when performing simple arithmetic operations, but it has a high communication overhead when processing complex arithmetic operations (as can be seen later on the performance comparison of S2PLoR). GC requires multiple encryption and decryption of gate circuits when dealing with arithmetic operations, which affects arithmetic performance. OT has a poor overall performance due to the encryption operations and high communication overhead, while HE also performs poorly due to its computation and communication performed on encrypted data. In S2PS, we utilize the MASCOT \cite{Keller_Orsini_Scholl_2016} scheme implemented in MP-SPDZ for the computation, which improves security while greatly degrading the computational performance due to its exploitation of OT under the malicious model. EVA-S2PLoR ranks second only to SS in terms of runtime under LAN and WAN, and despite not dominating in terms of communication volume, it performs better in aggregate thanks to fewer fixed communication rounds and a concise computational process.

Fig. \ref{fig: Proportion of Verification Cost in S2PS and S2PHM} tells that as the vector length or matrix dimension increases, the proportion of verification shows a decreasing posture regardless of the value of the verification rounds $L$. At $L=20$, the proportion of verification on large dimensional data is very small ($9.8\%$ in S2PS and $13.9\%$ in S2PHM), indicating that the additional overhead incurred by verification is acceptable. And the verification failure rate at $L=20$ is extremely low ($7.52\times 10^{-37}$ in S2PS and $8.27\times 10^{-25}$ in S2PHM), which is fully sufficient in practical applications.

\subsubsection{Precision Evaluation}
\begin{figure}[htb]
    \centering
    \subfigure[MRE of EVA-S2PLoR]{\includegraphics[width=0.325\linewidth]{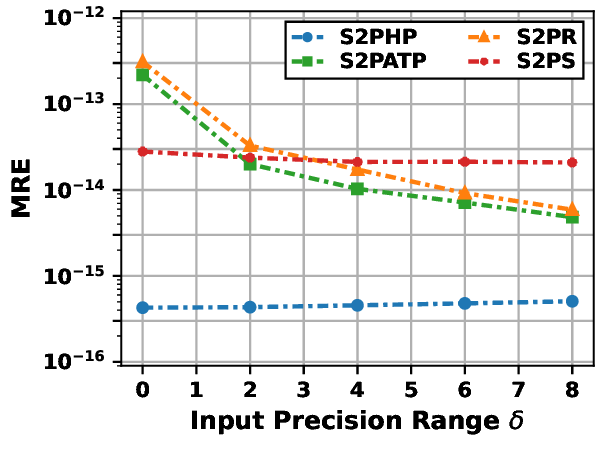}
    \label{fig: MRE of our Framework}}
    \hspace{-0.3cm}
    \subfigure[ARE of S2PHP]{\includegraphics[width=0.325\linewidth]{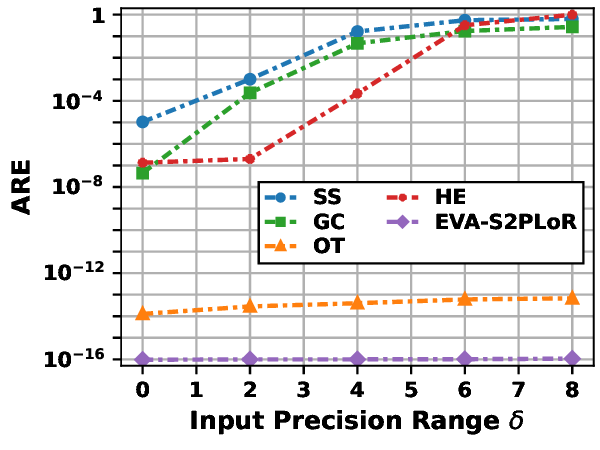}
    \label{fig: ARE Comparison of S2PHP}}
    \hspace{-0.3cm}
    \subfigure[ARE of S2PS]{\includegraphics[width=0.325\linewidth]{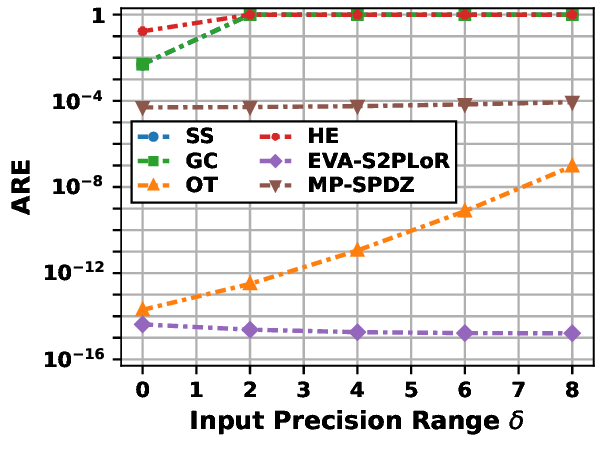}
    \label{fig: ARE Comparison of S2PS}}
    \caption{The maximum relative error (MRE) of our basic protocols in (a) and the average relative error (ARE) compared with five frameworks on S2PHP and S2PS protocols in (b)$\sim$(c) under five precision ranges $\delta$ with vector length of 500.}
    \label{fig: MRE and ARE comparison of Basic Protocols}
    \vspace{-0.2cm}
\end{figure}

We performed a precision evaluation on basic protocols in EVA-S2PLoR and compared it with other frameworks. The results are displayed in Fig. \ref{fig: MRE and ARE comparison of Basic Protocols}.

\textbf{Precision of EVA-S2PLoR:} In EVA-S2PLoR, operations on vectors can be viewed as parallel operations on numbers, so the analysis of precision can be explained from a numerical computation perspective. Fig. \ref{fig: MRE and ARE comparison of Basic Protocols}(a) shows the MRE of S2PHP increases slowly as the $\delta$ range increases, because the scale differences in the data increase as the $\delta$ range becomes larger, which introduces more round-off errors. However, there exists an upper bound to this increase, and since the numerical multiplication in S2PHP is implemented using vector dot products, according to \cite{Zarowski_2004}, we can get the upper error bound formula $MRE\le 1.25nu\frac{|\boldsymbol{x}|^T \cdot |\boldsymbol{y}|}{|{\boldsymbol {x}}^T\cdot \boldsymbol{y}|}$, where $n$ is the vector length ($n=m^2=4$ when the split number $m=2$), $u$ represents the floating-point precision (for 64-bit floats $u=2^{-52}$), and $\frac{|\boldsymbol{x}|^T \cdot |\boldsymbol{y}|}{|{\boldsymbol{x}}^T\cdot \boldsymbol{y}|} = 1$ when random splits are kept with the same sign in S2PHP, and therefore $MRE \le 1.25 \times 4 \times 2^{-52} \approx 1.11 \times 10^{-15}$. 
Note that the MRE of S2PATP and S2PR decreases as the $\delta$ range increases, due to the fact that when the $\delta$ range is small, there is a higher probability that the addition of signed random numbers will turn out to be a subtraction of similar numbers, leading to the loss of significant digits \cite{Sauer_2012}. Assuming that each digit and sign of the random numbers are generated independently, we can give the probability that at least $d$ digits are lost among n (vector length) repeated additions of random numbers as $Pr(n, d) = \frac{1-(1 - 10^{-d})^{n}}{2}$, where $Pr(500,3)\approx 19.7\%$ and $Pr(500,4)\approx 2.4\%$ is quite small. Since S2PATP and S2PR are based on S2PHP, we can roughly give the upper bound $MRE \le 1.11\times 10^{-15}\times10^{3} = 1.11\times 10^{-12}$.
The activation function $\sigma(x)$ can be regarded as 0 for $x\le -100$ and 1 for $x \ge 100$, so even if there is an error in $x$, it will barely affect the results. Therefore, the S2PS shows a higher probability of not generating errors as the $\delta$ range increases, resulting in a slow decreasing trend of its MRE. Furthermore, since S2PS does not suffer from the subtraction of similar numbers, its MRE is lower than that of S2PATP and S2PR when the $\delta$ range is small.

\textbf{Precision Comparison}: Fig. \ref{fig: MRE and ARE comparison of Basic Protocols}(b)$\sim$(c) illustrate that EVA-S2PLoR has the best precision on S2PHP and S2PS, and OT is in the second place due to the fact that OT primarily protects the transmission process and only introduces errors during data splitting, so the errors are relatively small. In S2PHP, HE, due to its operation in the ring $\mathbb{Z}_{2^k}$, demonstrates stable precision when the $\delta$ range is small. However, its error increases rapidly once the range exceeds the precision representation limits. GC generates large errors when converting between arithmetic operations and boolean circuits, while SS also experiences significant precision loss due to fixed-point number operations. In S2PS, MP-SPDZ uses bit decomposition to convert exponential operations into partial integer and fractional operations, where the latter is then processed using a polynomial approximation, resulting in relatively stable precision performance. On the other hand, HE, GC, and SS use polynomials or segmented function approximations to realize the activation function, which results in a significant loss of precision. In general, through precision comparison experiments, it can be concluded that EVA-S2PLoR has an extremely large precision advantage over the existing mainstream frameworks.

\subsection{Performance of S2PLoR}
To validate the performance of EVA-S2PLoR, we trained and predicted a vertically partitioned 2-party logistic regression model on three common datasets, comparing performance with three state-of-the-art frameworks (SecretFlow, CrypTen and FATE) in terms of time and several evaluation metrics. Besides, we provided a plaintext model (PlainLoR) implemented in Python as a performance reference, serving as a standard to uniformly determine the model parameters.

\subsubsection{Datasets and Settings}
We used two small datasets \textbf{Raisin} \cite{raisin_850}, \textbf{German Credit} \cite{statlog_(german_credit_data)_144} and a large dataset \textbf{Mnist} \cite{Mnist_dataset} for our experiments (Details are shown in Appendix \ref{append: datasets}). 

\subsubsection{Metrics of Model Evaluation}
In the evaluation, we used five model metrics: Accuracy, Precision, Recall, F1-Score, and AUC, which are described in detail in the Appendix \ref{append: datasets}.


\subsubsection{Efficiency and Accuracy Analysis}
\begin{figure}[htb]
    \centering
    \subfigure[S2PLoRT Stage Time]{\includegraphics[width=0.45\linewidth]{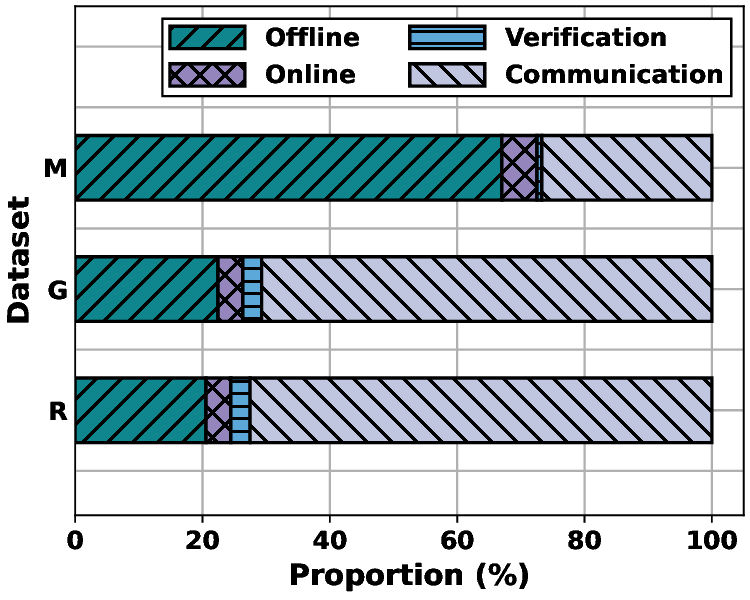}
    \label{fig: S2PLoRT Stage Time}}
    \hspace{0.3cm}
    \subfigure[S2PLoRP Stage Time]{\includegraphics[width=0.45\linewidth]{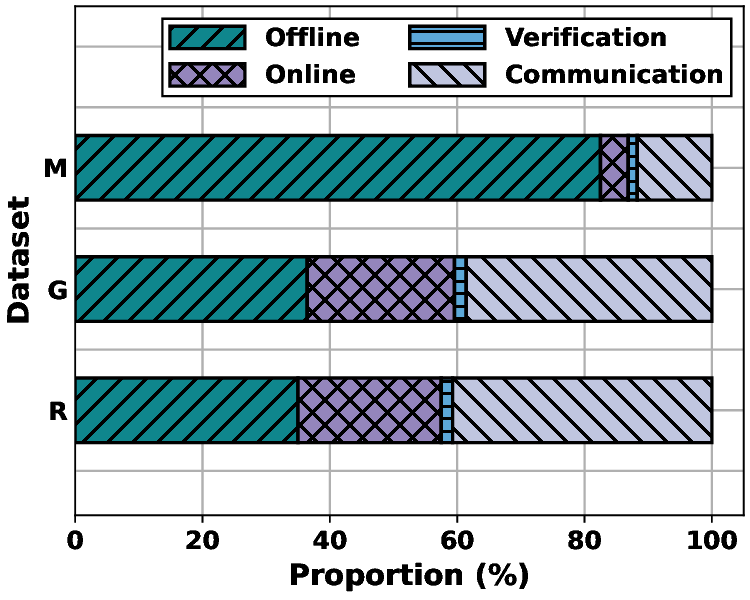}
    \label{fig: S2PLoRP Stage Time}}
    \caption{The allocation of runtime spent on different stages in S2PLoRT (a) and S2PLoRP (b), including offline, online, verification and communication costs, where we use three datasets: Raisin (R), German Credit (G) and Mnist (M).}
    \label{fig: EVA-S2PLoR Stage Time}
    \vspace{-0.2cm}
\end{figure}

We compared EVA-S2PLoR with three representative frameworks on three datasets where results of efficiency are demonstrated in Table \ref{tab:S2PLoR-Time}, and the model metrics are demonstrated in Table \ref{tab:S2PLoR-Accuracy}. Fig. \ref{fig: EVA-S2PLoR Stage Time} illustrates the allocation of time spent on different stages. S2PLoRP is computationally optimized on the \textbf{Mnist} dataset by predicting parallel in batches.

\begin{table}[htbp]
  \centering
  \caption{Efficiency comparison of S2PLoR under LAN and WAN with SecretFlow, CrypTen, and Fate on the Raisin, German Credit and Mnist where PlainLoR serves as a reference.}
  \resizebox{\linewidth}{!}{
    \begin{tabular}{cccccccccccccccc}
    \toprule  
    \multicolumn{1}{c}{\multirow{2}[2]{*}{\textbf{Dataset}}} & 
    \multicolumn{1}{c}{\multirow{2}[2]{*}{\textbf{Framework}}} & 
    \multicolumn{2}{c}{\textbf{Training Overhead}} & & 
    \multicolumn{2}{c}{\textbf{Training Time (s)}} & & 
    \multicolumn{2}{c}{\textbf{Predicting Overhead}} & & 
    \multicolumn{2}{c}{\textbf{Predicting Time (s)}} \\
    \cmidrule{3-4} \cmidrule{6-7} \cmidrule{9-10} \cmidrule{12-13}
    &  & \textbf{Comm. (MB)} & \textbf{Rounds} & & \textbf{LAN} & 
    \textbf{WAN} & & \textbf{Comm. (MB)} & \textbf{Rounds} & & \textbf{LAN} & \textbf{WAN} \\
    \midrule
    \multirow{4.5}[2]{*}{\textbf{Raisin}} & 
    \textbf{SecretFlow} & 11.17 & 8731 & & 6.04 & 354.70 & & 1.59 & 562 & & 1.24 & 23.70 \\
    & \textbf{CrypTen} & 66.29 & 69000 & & 15.20 & 2770.01 & & 0.27 & 152 & & 0.03 & 6.10 \\
    & \textbf{Fate} & 3.07 & 1361 & & 254.66 & 309.04 & & 0.81 & 511 & & 29.77 & 50.18 \\
    & \textbf{EVA-S2PLoR} & 22.73 & 4025 & & \textbf{0.58} & \textbf{161.77} & & 5.33 & 25 & & \textbf{0.02} & \textbf{1.15} \\
    & \textbf{PlainLoR} & - & - & & \textbf{1.02E-03} & - & & - & - & & \textbf{6.44E-06} & - \\
    \midrule
    \multirow{4.5}[2]{*}{\makecell{\textbf{German}\\\textbf{Credit}}} & 
    \textbf{SecretFlow} & 13.02 & 9779 & & 9.26 & 399.78 & & 2.84 & 647 & & 1.29 & 27.18 \\
    & \textbf{CrypTen} & 74.86 & 75000 & & 15.91 & 3010.35 & & 0.32 & 152 & & 0.03 & 6.10 \\
    & \textbf{Fate} & 3.53 & 1493 & & 289.56 & 349.22 & & 0.94 & 556 & & 29.80 & 52.01 \\
    & \textbf{EVA-S2PLoR} & 28.86 & 4375 & & \textbf{0.65} & \textbf{175.96} & & 6.64 & 25 & & \textbf{0.02} & \textbf{1.19} \\
    & \textbf{PlainLoR} & - & - & & \textbf{1.08E-03} & - & & - & - & & \textbf{6.20E-06} & - \\
    \midrule
    \multirow{4.5}[2]{*}{\textbf{Mnist}} & 
    \textbf{SecretFlow} & 1662.91 & 81850 & & 83.53 & 3392.40 & & 252.26 & 3560 & & 1.77 & 150.34 \\
    & \textbf{CrypTen} & 3602.60 & 547829 & & 132.14 & 22083.80 & & 71.32 & 667 & & \textbf{0.74} & 29.20 \\
    & \textbf{Fate} & 478.76 & 45585 & & 10777.51 & 12608.75 & & 62.83 & 1247 & & 37.13 & 88.51 \\
    & \textbf{EVA-S2PLoR} & 8383.76 & 32830 & & \textbf{36.58} & \textbf{1563.58} & & 254.50 & 25 & & 1.69 & \textbf{9.28} \\
    & \textbf{PlainLoR} & - & - & & \textbf{1.88E-01} & - & & - & - & & \textbf{1.21E-03} & - \\
    \bottomrule
    \end{tabular}}%
  \label{tab:S2PLoR-Time}%
\end{table}%

\begin{table}[htbp]
  \centering
  \caption{Accuracy evaluation of S2PLoR compared with three mainstream frameworks.}
  \resizebox{\linewidth}{!}{
    \begin{tabular}{cccccccccc}
    \toprule  
    \multicolumn{1}{c}{\multirow{2}[2]{*}{\textbf{Dataset}}} & 
    \multicolumn{1}{c}{\multirow{2}[2]{*}{\textbf{Framework}}} & 
    \multicolumn{5}{c}{\textbf{Accuracy Evaluation}} \\
    \cmidrule{3-7} 
    & & \textbf{Accuracy} & \textbf{Precision} & \textbf{Recall} & 
    \textbf{F1-Score} & \textbf{AUC}  \\
    \midrule
    \multirow{5}[2]{*}{\textbf{Raisin}} & 
    \textbf{SecretFlow} & 0.8333 & 0.8488 & 0.8111 & 0.8295 & 0.8874 \\
    & \textbf{CrypTen} & 0.8389 & 0.8861 & 0.7778 & 0.8284 & 0.8875 \\
    & \textbf{Fate} & 0.8389 & 0.8588 & 0.8111 & 0.8343 & 0.8867 \\
    & \textbf{EVA-S2PLoR} & \textbf{0.8500} & 0.9091 & 0.7778 & \textbf{0.8383} & \textbf{0.8885} \\
    & \textbf{PlainLoR} & \textbf{0.8556} & 0.9103 & 0.7889 & \textbf{0.8452} & \textbf{0.8896} \\
    \midrule
    \multirow{5}[2]{*}{\makecell{\textbf{German}\\\textbf{Credit}}} & 
    \textbf{SecretFlow}  & 0.7900 & 0.6792 & 0.5902 & 0.6316 & 0.8155 \\
    & \textbf{CrypTen} & 0.7850 & 0.6957 & 0.5246 & 0.5981 & 0.8114 \\
    & \textbf{Fate} & 0.7100 & 0.5349 & 0.3770 & 0.4423 & 0.7496 \\
    & \textbf{EVA-S2PLoR} & \textbf{0.8050} & 0.6897 & 0.6557 & \textbf{0.6723} & \textbf{0.8175} \\
    & \textbf{PlainLoR} & \textbf{0.8100} & 0.6949 & 0.6721 & \textbf{0.6833} & \textbf{0.8184} \\
    \midrule
    \multirow{5}[2]{*}{\textbf{Mnist}} &
    \textbf{SecretFlow} & 0.9876 & 0.9957 & 0.9906 & 0.9931 & 0.9925 \\
    & \textbf{CrypTen} & 0.9902 & 0.9953 & 0.9938 & 0.9946 & 0.9963 \\
    & \textbf{Fate} & 0.8537 & 0.9745 & 0.8603 & 0.9139 & 0.8995 \\
    & \textbf{EVA-S2PLoR} & \textbf{0.9915} & 0.9964 & 0.9941 & \textbf{0.9953} & \textbf{0.9964} \\
    & \textbf{PlainLoR} & \textbf{0.9921} & 0.9956 & 0.9957 & \textbf{0.9956} & \textbf{0.9975} \\
    \bottomrule
    \end{tabular}}%
  \label{tab:S2PLoR-Accuracy}%
\end{table}%
As can be seen in Tables \ref{tab:S2PLoR-Time} and \ref{tab:S2PLoR-Accuracy}, EVA-S2PLoR achieves the best overall performance in terms of efficiency and model metrics. Fate, based on the federated learning paradigm, has lower communication overhead but suffers from extremely high computational costs due to its utilization of HE for privacy protection. Both SecretFlow and CrypTen utilize SS, resulting in a large number of communication rounds, which negatively affects the overall efficiency performance. Despite EVA-S2PLoR having a larger communication volume, it possesses a smaller time overhead due to fewer communication rounds and simpler computation. Failing to perform accurate computations on nonlinear operations in three frameworks makes them perform less well than EVA-S2PLoR on model metrics. In general, EVA-S2PLoR outperforms many advanced frameworks in privacy logistic regression tasks.

From Fig. \ref{fig: EVA-S2PLoR Stage Time}, it can be seen that the main components of model training and prediction time in the LAN environment are offline computation and communication. The offline time on the large dataset (\textbf{Mnist}) exceeds the communication time, due to the high bandwidth and low latency in the LAN environment. Note that the time spent on verification in the actual application of logistic regression is very small, reflecting the lightweight nature of our verification module.

\section{Related Work}\label{section 7: Related Work}
\subsection{Secure Vector Operations}
For SS based methods, SecureML\cite{Mohassel_Zhang_2017} utilizes OT or linearly homomorphic encryption to protect generated vectorized triplets, enabling secure 2-party vector inner product operations. ABY3\cite{Mohassel_Rindal_2018} employs shared truncation to perform secure 3-party vector inner product operations with truncation. BLAZE\cite{Patra_Suresh_2020} optimizes this truncation protocol to provide a more efficient secure vector inner product operation with truncation. FLASH\cite{Byali_Chaudhari_Patra_Suresh_2020} adopts a unique couple-sharing structure to address the issue that communication complexity is related to the vector size when performing inner product operations in malicious security models, which is faced by frameworks such as ABY3, SecureNN\cite{Wagh_Gupta_Chandran_2019}, and ASTRA\cite{Chaudhari_Choudhury_Patra_Suresh_2019}. Daalen\cite{vanDaalen_Ippel_Dekker_Bermejo_2023} uses DD to transform the vector inner product into diagonal matrix multiplication, enabling multi-party vector inner product operations. SWIFT\cite{Koti_Pancholi_Patra_Suresh_2021}, based on a divide-and-conquer approach, employs secure comparison protocols to implement secure vector maximum value operations. TenSEAL\cite{Benaissa_Retiat_Cebere_Belfedhal_2021}, built on the CKKS scheme of SEAL\cite{Cheon_Kim_Kim_Song_2017}, utilizes HE to perform vector inner product and hadamard product operations.

\subsection{Secure Logistic Regression}
In earlier years, Fienberg\cite{Fienberg_Nardi_Slavković_2009} proposed a secure maximum likelihood estimation method for log-linear models based on secure summation, introducing a secure logistic regression algorithm for horizontally partitioned data. Subsequently, Slavkovic\cite{Slavkovic_Nardi_Tibbits_2007} employed the Newton-Raphson method and SMPC techniques to develop secure logistic regression algorithms for both horizontally and vertically partitioned data, though these lacked rigorous security proofs. Chaudhuri\cite{Chaudhuri_Monteleoni_2008}, using the $\epsilon$-differential privacy framework\cite{Dwork_McSherry_Nissim_Smith_2006}, implemented two privacy-preserving logistic regression classifiers and established the relationship between the regularization constant and the amount of differential privacy noise added to protect logistic regression privacy. Duverle\cite{Duverle_Kawasaki_Yamada_Sakuma_Tsuda_2015} proposed a sampling-based secure protocol and implemented a secure and accurate logistic regression method using HE, although its applications are limited to statistical analysis. Du\cite{Du_Li_Li_2018} employed HE and an asynchronous gradient sharing protocol to achieve secure 2-party logistic regression under vertically partitioned data without a third-party computational node.

In recent years, since SecureML integrated multiple privacy-preserving computation techniques to form a privacy machine learning framework, hybrid protocol frameworks such as EzPC\cite{Chandran_Gupta_Rastogi_Sharma_Tripathi_2019}, PrivColl\cite{Zhang_Bai_Li_Curtis_Chen_Ko_2020}, Trident\cite{Rachuri_Suresh_Chaudhari_2020}, MP-SPDZ\cite{Keller_2020}, CrypTen\cite{Knott_Venkataraman_Hannun_Sengupta_Ibrahim_vanderMaaten_2021}, and SecretFlow\cite{Ma_Zheng_Feng_Zhao_Wu_Fang_Tan_Yu_Zhang_Wang_2023} have emerged, supporting not only the privacy-preserving logistic regression but also more complex machine learning models.


\section{Conclusion and Discussion}\label{section 8: Conclusion and Discussion}


This paper proposed an efficient, verifiable, and accurate secure hadamard product protocol and its derived protocols. A high-precision secure protocol for the sigmoid function $\sigma(x)$ was designed for logistic regression and other decentralized PPML scenarios. The efficiency and precision of the basic operators are verified through experiments, and secure 2-party logistic regression experiments are conducted on three datasets, where EVA-S2PLoR achieves the best overall performance in comparison with three state-of-the-art frameworks.

Introducing the split factor $\rho$ for secure verification increases computational complexity, resulting in greater overhead. Therefore, designing more efficient verification methods will be a focus of our future work. In this paper, we set $\rho=2$ to achieve efficient computation. Furthermore, $\rho$ is a key indicator of data leakage risk. A smaller $\rho$ suggests poorer data dispersion, increasing the risk of adversaries reconstructing original information. In future work, we will analyze the impact of $\rho$ on protocol security and efficiency.
Finally, this paper primarily focuses on the exposure of data boundaries. Future research will further explore the privacy risks associated with the statistical information of input variables.

\appendices

\section{Proof of verifiability}\label{append: verification proof}
We give the proof of the following Theorem \ref{theorem: S2PM verification}.
\begin{theorem}\label{theorem: S2PM verification}
The verification failure rate of S2PM does not exceed $4^{-l}$, where $l$ represents the verification rounds.
\end{theorem}


\begin{IEEEproof}
    When the computation is correct, i.e., $V_a + V_b = A \times B$, $E_r$ will always be \textbf{0} regardless of $\hat{\delta_a}$. Because according to Algorithm \ref{alg:S2PM-Computing}, we have:
\begin{flalign}
     VF_a + VF_b - S_t &= V_a + V_b + R_a \times \hat{B} - \hat{A} \times B - S_t \nonumber \\
                       &= (V_a + V_b - A \times B) + R_a \times R_b - S_t \nonumber \\
                       &= V_a + V_b - A \times B &
\end{flalign}

\noindent When the computation is abnormal, i.e., $V_a + V_b \neq A \times B$. Let $P_f(l)$ denote the probability that Alice fails to detect the error after $l$ rounds of verification. We will show that $P_f(l)$ is upper-bounded by the exponential of $l$ and is negligible with sufficiently large $l$. 
Let $H = VF_a + VF_b - S_t$, and $E_r = H \times \hat{\delta_a} = (p_1, \cdots, p_n)^T$. There should be at least one non-zero element in $H$, say $h_{ij} \neq 0$ and its corresponding $p_i$ in $E_r$, with the relationship $p_i = h_{ij} \delta_j + w$, where $w = \sum_{k=1}^m h_{ik} \delta_k - h_{ij} \delta_j$. The marginal probability $P(p_i = 0)$ can be calculated as:
\begin{align}
    P(p_i = 0) = &P(p_i = 0 | w = 0)\cdot P(w = 0) \; + \nonumber \\
    &P(p_i = 0 | w \neq 0)\cdot P(w \neq 0)
\end{align}
Substituting the posteriors
\begin{align}
P(p_i = 0 | w = 0) = P(\delta_j = 0) = 1/2 \nonumber \\
P(p_i = 0 | w \neq 0) \leq P(\delta_j = 1) = 1/2      
\end{align}
gives
\begin{equation}
    P(p_i = 0) \leq (1/2)\cdot P(w = 0) + (1/2)\cdot P(w \neq 0)    
\end{equation}
Because $P(w \neq 0) = 1 - P(w = 0)$, we have
\begin{equation}
    P(p_i = 0) \leq 1/2
\end{equation}
The probability of single check failure $P_f(1)$ satisfies
\begin{equation}
    P_f(1) = P(E_r = (0, \cdots, 0)^T) \leq P(p_i = 0) \leq 1/2
\end{equation}
Since each round of verification is independent, $P_f(l)$ satisfies
\begin{equation}
    P_f(l) = P_f(1)^l \leq 2^{-l}
\end{equation}
Verification fails if and only if both parties fail, so we have:

\begin{equation}
    P_f(S2PM) = (P_f(l))^2 \leq 4^{-l}
\end{equation}
The proof is now completed.
\end{IEEEproof}

\section{Proof of Security}\label{append: security proof}
We first provide definitions of indistinguishability security.

\begin{definition}[Computational Indistinguishability \cite{Goldreich_2004}]\label{def2}
     A probability ensemble $X = \{X(a, n)\}_{a\in\{0,1\}^{*};n\in \mathbb{N}}$ is an infinite sequence of random variables indexed by $a\in \{0, 1\}^{*}$ and $n\in \mathbb{N}$. In the context of secure computation, $a$ represents the input of the parties and $n$ represents the security parameter. Two probability ensembles $X = \{X(a, n)\}_{a\in\{0,1\}^{*};n\in \mathbb{N}}$ and $Y = \{Y(a, n)\}_{a\in \{0,1\}^{*};n\in \mathbb{N}}$ are said to be computationally indistinguishable, denoted by $X\overset{c}{\equiv}Y$, if for every nonuniform polynomial-time algorithm $D$ there exists a negligible function $\mu(\cdot)$ such that for every $a\in \{0, 1\}^{*}$ and every $n\in \mathbb{N}$,
\begin{align}
    |Pr[D(X(a,n))=1]-Pr[D(Y(a,n))=1]|\leq \mu(n)  
\end{align}
\end{definition}

\begin{definition}[Security in Semi-honest 2-Party Computation \cite{Lindell_2017}]\label{def3}
Let $f:\{0,1\}^{*}\times \{0,1\}^{*}\mapsto \{0,1\}^{*}\times\{0,1\}^{*}$ be a probabilistic polynomial-time functionality, where $f=(f_1(x,y),f_2(x,y))$ and $\pi$ be a 2-party protocol for computing $f$. The view of the $i$-th party ($i\in \{1,2\}$) during an execution of $\pi$ on $(x,y)$ is denoted by $view^{\pi}_{i}(x, y)$ and equals $(w,r^i;m^i_1,\cdots,m^i_t)$, where $w\in\{x,y\}$ (its input depending on the value of $i$), $r^i$ represents the contents of the $i$-th party’s internal random tape, and $m^i_j$ represents the $j$-th message the $i$-th party received. The output of the $i$-th party during an execution of $\pi$ on $(x,y)$, denoted by $output^{\pi}_i(x,y)$, can be computed from its own view of the execution. We say that $\pi$ securely computes $f$ if there exist probabilistic polynomial-time algorithms $S_1$ and $S_2$ such that:
\begin{align}
 \{(S_1(x,f_1),f_2)\}
 \overset{c}{\equiv} \{(view^{\pi}_1(x, y), output^{\pi}_2(x, y))\} \nonumber \\
 \{(f_1, S_2(y, f_2))\}
 \overset{c}{\equiv} \{(output^{\pi}_1(x, y), view^{\pi}_2(x, y))\}
 \end{align}
 where $x,y\in \{0,1\}^{*}$ such that $|x|=|y|$.
\end{definition}

According to the definition of security in semi-honest 2-party computation above, let $f=(f_1,f_2)$ be a probabilistic polynomial-time function, and let $\pi$ be a 2-party protocol for computing $f$. We say $\pi$ securely computes $f$ if we can construct two simulators in the ideal-world ($S_1$ and $S_2$), such that the following relations hold at the same time:
\begin{align}
    \{S_1(x, f_1(x,y))\}_{x,y\in \{0,1\}^{*}} \overset{c}{\equiv} \{view_1^{\pi}(x,y)\}_{x,y\in \{0,1\}^{*}} \nonumber \\
    \{S_2(y, f_2(x,y))\}_{x,y\in \{0,1\}^{*}} \overset{c}{\equiv} \{view_2^{\pi}(x,y)\}_{x,y\in \{0,1\}^{*}} \label{12}
\end{align}
and the output of one party remains unchanged regardless of the input from another party:
\begin{align}
    f_1(x,y) \equiv f_1(x,y^*) \nonumber \\
    f_2(x,y) \equiv f_2(x^*,y)\label{13}
\end{align}
In this way, we can transform the security proof problem into a construction problem. We use the lemma \ref{lemma1} below in the process of proof.

\begin{lemma}\label{lemma1}
For a linear system $A\cdot X=B$, if $rank(A)=rank(A|B)<n$ (where n is the number of rows of the matrix $X$ and $A|B$ is the augmented matrix), then this linear system has infinite solutions \cite{Shores_2018}.
\end{lemma}

We provide the security proof of S2PHP under the semi-honest adversary model below. In this proof, we denote the process of transforming vectors $\boldsymbol{a}, \boldsymbol{b}$ into matrices $A, B$ in algorithm \ref{alg:S2PHP} by $A = Ra2A(\boldsymbol{a},\rho)$ and $B = Rb2B(\boldsymbol{b},\rho)$, respectively. We also use $A = v2diag(\boldsymbol{a})$ to denote the generation of a matrix $A$ whose main diagonal is sequentially formed by elements in a vector $\boldsymbol{a}$, and the remaining elements of $A$ are generated randomly.

\begin{IEEEproof}
    We will construct two simulators $S_1$ and $S_2$.
    
    \textbf{Corrupted Alice:} Assuming Alice is the adversary, this means we have to construct a simulator $S_1$ simulating ${view}_1^\pi (x,y)=(\boldsymbol{a},\rho,R_a, r_a, S_t, r; \hat{B}, T, VF_b)$ such that $S_1(x, f_1(x,y))$ is indistinguishable from ${view}_1^\pi (x,y)$. Formally, $S_1$ receives ($\boldsymbol{a}$, $\rho$, $R_a$, $r_a$, $S_t$, $\boldsymbol{v_a}$) and a random tape $r$, and proceeds as follows:
    
    \begin{enumerate}
        \item $S_1$ uses the random tape $r$ to generate $A = Ra2A( \boldsymbol{a}, \rho)$, and computes $\hat{A} = A + R_a$ and $r_b = S_t - r_a$.
        \item $S_1$ generates two random vectors $\boldsymbol{b}'$ and $\boldsymbol{v_b}'$ such that $\boldsymbol{a} \odot \boldsymbol{b}' = \boldsymbol{v_a} + \boldsymbol{v_b}'$, and then generates $B' = Rb2B(\boldsymbol{b}', \rho)$ and $V_b' = v2diag(\boldsymbol{v_b}')$. \label{proof1_step2}
        \item $S_1$ generates a random matrix $R_b'$ where $R_a \times R_b'=S_t$, and computes $\hat{B}' = B' + R_b'$, $T' = \hat{A} \times B' + r_b - V_b'$ and $VF_b' = V_b' - \hat{A} \times B'$. \label{proof1_step3}
        \item Get $S_1(x, f_1(x,y)) = (\boldsymbol{a},\rho,R_a, r_a, S_t, r; \hat{B}', T', VF_b')$.
    \end{enumerate}
    In step \ref{proof1_step2}, we observe that there are infinite pairs of $(\boldsymbol{b}', \boldsymbol{v_b}')$ satisfying the equation $\boldsymbol{a} \odot \boldsymbol{b}' = \boldsymbol{v_a} + \boldsymbol{v_b}'$, so that $\boldsymbol{b}'$ and $\boldsymbol{v_b}'$ are simulatable, denoted by $\{\boldsymbol{b}', \boldsymbol{v_b}'\} \overset{c}{\equiv} \{\boldsymbol{b}, \boldsymbol{v_b} \}$. In this case, we can figure out that $B'$ and $V_b'$ are simulatable as well, since $B' = Rb2B(\boldsymbol{b}', \rho)$ and $V_b' = v2diag(\boldsymbol{v_b}')$. In step \ref{proof1_step3}, according to the lemma \ref{lemma1}, we learn that there are infinite number of $R_b'$ where $R_a \times R_b'=S_t$ when $R_a$ is a rank-deficient matrix, and therefore $R_b'$ is simulatable. Considering that the variables $\hat{B}', T', VF_b'$ are computed from the simulatable variables $B', R_b', V_b'$, we can easily prove that they are all simulatable. Thus, we obtain that $\{(\boldsymbol{a},\rho,R_a, r_a, S_t, r; \hat{B}', T', VF_b')\} \overset{c}{\equiv} \{(\boldsymbol{a},\rho,R_a, r_a, S_t, r; \hat{B}, T, VF_b) \}$, i.e., $\{S_1(x, f_1)\}_{x,y} \overset{c}{\equiv} \{view_1^{\pi}(x,y)\}_{x,y}$. Besides, $V_a' = T' + r_a - R_a \cdot \hat{B}' = \hat{A} \times {B}' - R_a \times \hat{B}' + r_b + r_a - {V_b}' = A \times \hat{B}' - {V_b}'$, so that $\boldsymbol{v_a}' = diag2v(V_a') = diag2v(A \times \hat{B}' - {V_b}')$. According to the correctness of S2PHP, we can conclude that $\boldsymbol{v_a}' = \boldsymbol{v_a}$, which shows that for any arbitrary input of Bob, the output of Alice $\boldsymbol{v_a}$ remains unchanged, i.e., $f_1(x,y) \equiv f_1(x,y^*)$.\\

    \textbf{Corrupted Bob:} Assuming Bob is the adversary, this means we have to construct a simulator $S_2$ simulating ${view}_2^\pi (x,y)=(\boldsymbol{b},\rho,R_b, r_b, S_t, r_0, r_1; \hat{A}, VF_a)$ such that $S_2(y, f_2(x,y))$ is indistinguishable from ${view}_2^\pi (x,y)$. Formally, $S_2$ receives $(\boldsymbol{b},\rho,R_b, r_b, S_t, \boldsymbol{v_b})$ and two random tapes $r_0, r_1$, and proceeds as follows:
    
    \begin{enumerate}
        \item $S_2$ uses the random tape $r_0$ to generate $B = Rb2B( \boldsymbol{b}, \rho)$, and computes $\hat{B} = B + R_b$. Meanwhile, $S_2$ uses the random tape $r_1$ to generate matrix $V_b$ where $\boldsymbol{v_b} = diag2v(V_b)$.
        \item $S_2$ generates two random vectors $\boldsymbol{a}'$ and $\boldsymbol{v_a}'$ such that $\boldsymbol{a}' \odot \boldsymbol{b} = \boldsymbol{v_a}' + \boldsymbol{v_b}$, and then generates $A' = Ra2A(\boldsymbol{a}', \rho)$ and $V_a' = v2diag(\boldsymbol{v_a}')$. \label{proof2_step2}
        \item $S_2$ generates a random matrix $R_a'$ where $R_a' \times R_b = S_t$, and computes $\hat{A}' = A' + R_a'$ and $VF_a' = V_a' + R_a' \times \hat{B}$. \label{proof2_step3}
        \item Get $S_2(y, f_2(x,y)) = (\boldsymbol{b},\rho,R_b, r_b, S_t, r_0, r_1; \hat{A}', VF_a')$.
    \end{enumerate}
    In step \ref{proof2_step2}, we observe that there are infinite pairs of $(\boldsymbol{a}', \boldsymbol{v_a}')$ satisfying the equation $\boldsymbol{a}' \odot \boldsymbol{b} = \boldsymbol{v_a}' + \boldsymbol{v_b}$, so that $\boldsymbol{a}'$ and $\boldsymbol{v_a}'$ are simulatable, denoted by $\{\boldsymbol{a}', \boldsymbol{v_a}'\} \overset{c}{\equiv} \{\boldsymbol{a}, \boldsymbol{v_a} \}$. In this case, we can figure out that $A'$ and $V_a'$ are simulatable as well, since $A' = Ra2A(\boldsymbol{a}', \rho)$ and $V_a' = v2diag(\boldsymbol{v_a}')$. In step \ref{proof2_step3}, according to the lemma \ref{lemma1}, we learn that there are infinite number of $R_a'$ where $R_a'\times R_b=S_t$ when $R_b$ is a rank-deficient matrix, and therefore $R_a'$ is simulatable. Considering that the variables $\hat{A}', VF_a'$ are computed from the simulatable variables $A', R_a', V_a'$, we can easily prove that they are all simulatable. Thus, we obtain that $\{(\boldsymbol{b},\rho,R_b, r_b, S_t, r_0, r_1; \hat{A}', VF_a')\} \overset{c}{\equiv} \{(\boldsymbol{b},\rho,R_b, r_b, S_t, r_0, r_1; \hat{A}, VF_a)\}$, i.e., $\{S_2(y, f_2)\}_{x,y} \overset{c}{\equiv} \{view_2^{\pi}(x,y)\}_{x,y}$. Besides, due to $\boldsymbol{v_b}' = diag2v(V_b) = \boldsymbol{v_b}$, it is shown that for any arbitrary input of Alice, the output of Bob $\boldsymbol{v_b}$ remains unchanged, i.e., $f_2(x,y) \equiv f_2(x^*,y)$.
\end{IEEEproof}

The universally composable security theory (UC) \cite{Canetti_2001} promises the security of a protocol when all its sub-protocols are secure. Therefore, the security of other protocols in EVA-S2PLoR is easily proved by the following lemma \cite{Bogdanov_Laur_Willemson_2008}. 

\begin{lemma}\label{lemma2}
    A protocol is perfectly simulatable if all sub-protocols are perfectly simulatable.
\end{lemma}


\section{Datasets and Metrics in Evaluation of S2PLoR}\label{append: datasets}

\subsection{Metrics of Model Evaluation}
When using a logistic regression model for a binary classification task, the predicted values of the model $\hat{y} \in [0, 1]$ are continuous values, and the samples are labelled $y \in \{0, 1\}$ as either a positive class (denoted as 1) or a negative class (denoted as 0), and therefore a threshold $\theta \in [0, 1]$ is usually set up when performing the model evaluation to convert the predicted value $\hat{ y}$ into 0 (if $\hat{y} < \theta$) and 1 (if $\hat{y} \ge \theta$).

At a given threshold, four statistical metrics can be derived: True Positive (TP), False Positive (FP), True Negative (TN), and False Negative (FN). Here, TP (resp., TN) represents the number of test cases where the model correctly predicts the positive (resp., negative) class. Conversely, FP (resp., FN) refers to the number of negative (resp., positive) instances where the model incorrectly predicts the positive (resp., negative) class. Using these metrics, we can derive several commonly used evaluation metrics, including Accuracy = $\frac{TP+TN}{TP+TN+FP+FN}$ (overall correctness), Precision = $\frac{TP}{TP + FP}$ (positive prediction accuracy), Recall = $\frac{TP}{TP + FN}$ (positive detection ability), F1-Score = $2\cdot \frac{Precision \cdot Recall}{Precision + Recall}$ (Precision-Recall balance), True positive rate (TPR, equivalent to Recall) and False positive rate (FPR = $\frac{FP}{FP+TN}$, rate of false alarms). Furthermore, by adjusting the threshold, we can obtain different TPR and FPR values. Plotting FPR on the horizontal axis and TPR on the vertical axis yields the Receiver Operating Characteristic (ROC) curve. The area under the curve (AUC) serves as a key metric to quantify the overall classification performance.

\subsection{Datasets and Settings}
Details of the datasets used in S2PLoR evaluation are shown in Table \ref{tab:Datasets}. Both \textbf{Raisin} and \textbf{German Credit} are binary datasets that can be directly used for logistic regression, while for the multi-class dataset \textbf{Mnist}, we simplified its labels into two classes: zero and non-zero.

\begin{table}[h]
  \centering
  \caption{Dataset Details in S2PLoR}
    \resizebox{\linewidth}{!}{
    \begin{tabular}{ccccccc}
    \toprule
    \textbf{Dataset} & 
    \textbf{Features} &
    \textbf{Training Cases} &
    \textbf{Test Cases} &
    \textbf{Batch Size} &
    \textbf{Learning Rate} &
    \textbf{Iterations}
    \\
    \midrule
   \textbf{Raisin} & 7 &  720 & 180 & 32 & 0.05 & 5 \\
   \textbf{German Credit} & 20 &  800 & 200 & 32 & 0.05 & 5 \\
   \textbf{Mnist} & 784 &  60000 & 10000 & 128 & 0.25 & 2 \\
    \bottomrule
    \end{tabular}}%
  \label{tab:Datasets}%
\end{table}%

For secure logistic regression training and prediction involving Alice and Bob, each dataset is vertically divided such that features are evenly distributed between Alice and Bob as private data, while the labels are treated as public values.

\section*{Acknowledgment}

This research was supported by the Science and Technology Innovation Program of Xiongan New Area (Grant No. 2022XAGG0148) and the National Key Research and Development Program of China (Grant No. 2021YFB2700300).


\bibliographystyle{IEEEtran}
\bibliography{IEEEabrv,biblio}

\end{document}